\documentclass{aa}
\def\simlt{\lower.5ex\hbox{$\; \buildrel < \over \sim \;$}}
\usepackage{graphicx}
\begin{document}
\title{The VIMOS VLT Deep Survey
         \thanks{based on data
         obtained with the European Southern Observatory Very Large
         Telescope, Paranal, Chile, program 070.A-9007(A), and on data
         obtained at the Canada-France-Hawaii Telescope, operated by
         the CNRS of France, CNRC in Canada and the University of Hawaii}
}
\subtitle{Evolution of the luminosity functions by galaxy type up
to z=1.5 
\\
from first epoch data
}

   \author{ E.Zucca       \inst{1} 
   \and     O.Ilbert      \inst{2,3}
   \and     S.Bardelli    \inst{1}
   \and     L.Tresse      \inst{3}
   \and     G.Zamorani    \inst{1} 
   \and     S.Arnouts     \inst{3}
   \and     L.Pozzetti    \inst{1} 
   \and     M.Bolzonella  \inst{2} 
   \and     D.Bottini     \inst{4}
   \and     B.Garilli     \inst{4}
   \and     V.Le~Brun     \inst{3}
   \and     O.Le~F\`evre  \inst{3}
   \and     D.Maccagni    \inst{4}
   \and     J.P.Picat     \inst{5}
   \and     R.Scaramella  \inst{6}
   \and     M.Scodeggio   \inst{4}
   \and     G.Vettolani   \inst{6}
   \and     A.Zanichelli  \inst{6}
   \and     C.Adami       \inst{3}
   \and     M.Arnaboldi   \inst{7}
   \and     A.Cappi       \inst{1}
   \and     S.Charlot     \inst{8,10}
   \and     P.Ciliegi     \inst{1}  
   \and     T.Contini     \inst{5}
   \and     S.Foucaud     \inst{4}
   \and     P.Franzetti   \inst{4}
   \and     I.Gavignaud   \inst{5,12}
   \and     L.Guzzo       \inst{9}
   \and     A.Iovino      \inst{9}
   \and     H.J.McCracken \inst{10,11}
   \and     B.Marano      \inst{2}  
   \and     C.Marinoni    \inst{9}
   \and     A.Mazure      \inst{3}
   \and     B.Meneux      \inst{3}
   \and     R.Merighi     \inst{1} 
   \and     S.Paltani     \inst{3}
   \and     R.Pell\`o     \inst{5}
   \and     A.Pollo       \inst{9}
   \and     M.Radovich    \inst{7}
   \and     M.Bondi       \inst{6}
   \and     A.Bongiorno   \inst{2}
   \and     G.Busarello   \inst{7}
   \and     O.Cucciati    \inst{9}
   \and     L.Gregorini   \inst{6}
   \and     F.Lamareille  \inst{5}
   \and     G.Mathez      \inst{5}
   \and     Y.Mellier     \inst{10,11}
   \and     P.Merluzzi    \inst{7}
   \and     V.Ripepi      \inst{7}
   \and     D.Rizzo       \inst{5}
          }

   \offprints{ Elena Zucca \email{elena.zucca@oabo.inaf.it} }

\institute{
INAF-Osservatorio Astronomico di Bologna,
via Ranzani 1, I-40127 Bologna (Italy)
\and
Universit\`a di Bologna, Dipartimento di Astronomia,
via Ranzani 1, I-40127 Bologna (Italy)
\and
Laboratoire d'Astrophysique de Marseille (UMR 6110), CNRS-Universit\'e de
Provence, BP8, F-13376 Marseille Cedex 12 (France)
\and
INAF-IASF, via Bassini 15, I-20133 Milano (Italy)
\and
Laboratoire d'Astrophysique de l'Observatoire Midi-Pyr\'en\'ees (UMR 5572),
14 avenue E. Belin, F-31400 Toulouse (France)
\and
INAF-IRA, Via Gobetti 101, I-40129 Bologna (Italy)
\and
INAF-Osservatorio Astronomico di Capodimonte, 
via Moiariello 16, I-80131 Napoli (Italy)
\and
Max Planck Institut fur Astrophysik, 
D-85741 Garching bei M\"unchen (Germany)
\and
INAF-Osservatorio Astronomico di Brera,
via Brera 28, I-20121 Milano (Italy)
\and
Institut d'Astrophysique de Paris (UMR 7095), 
98 bis Bvd Arago, F-75014 Paris (France)
\and
Observatoire de Paris, LERMA, 61 Avenue de l'Observatoire, 
F-75014 Paris (France)
\and
European Southern Observatory, Karl-Schwarzschild-Strasse 2, 
D-85748 Garching bei M\"unchen (Germany)
          }

\authorrunning {E.Zucca et al.}

\titlerunning {The VVDS: luminosity functions by type up to z~=~1.5}

\date{Received -- -- ----; accepted -- -- ----}

\abstract{ 
From the first epoch observations of the VIMOS VLT Deep Survey
(VVDS) up to $z=1.5$ we have derived luminosity functions of different
spectral type galaxies.
The VVDS data, covering $\sim 70\%$ of the life of the Universe, allow for 
the first time to study from the same sample and with good statistical accuracy 
the evolution of the luminosity functions by galaxy type in several rest 
frame bands from a purely magnitude selected sample. 
The magnitude limit of the VVDS ($I_{AB}=24$) is significantly fainter
than the limit of other complete spectroscopic surveys and allows the
determination of the faint end slope of the luminosity function with unprecedented
accuracy. 
Galaxies have been classified in four spectral classes, from early type 
to irregular galaxies, using their colours and redshift, and luminosity functions 
have been derived in the U, B, V, R and I rest frame bands  
for each type, in redshift bins from $z=0.05$ to $z=1.5$. 
In all the considered rest frame bands, we find a significant  
steepening of the luminosity function going from early to late types.
The characteristic luminosity $M^*$ of the Schechter function
is significantly fainter for late type galaxies and this difference increases
in the redder bands.
Within each of the galaxy spectral types we find a brightening of $M^*$ 
with increasing redshift, ranging from $\simlt 0.5$ mag for early type
galaxies to $\sim 1$ mag for the latest type galaxies, while the slope 
of the luminosity function of each spectral type 
is consistent with being constant with redshift. The luminosity function
of early type galaxies is consistent with passive evolution up
to $z\sim 1.1$, while the number of bright ($M_{B_{AB}}< -20$) 
early type galaxies has
decreased by $\sim 40\%$ from $z\sim 0.3$ to $z\sim 1.1$. 
We also find a strong evolution in the normalization of the luminosity
function of latest type galaxies, 
with an increase of more than a factor $2$ from $z\sim 0.3$ to $z\sim 1.3$: 
the density of bright ($M_{B_{AB}}< -20$) late type galaxies in the same
redshift range increases of a factor $\sim 6.6$. 
These results indicate a strong type-dependent evolution and identifies
the latest spectral types as responsible for most of the evolution of
the UV-optical luminosity function out to $z=1.5$.

         }

\maketitle

\begin{figure}
\centering
\includegraphics[width=\hsize]{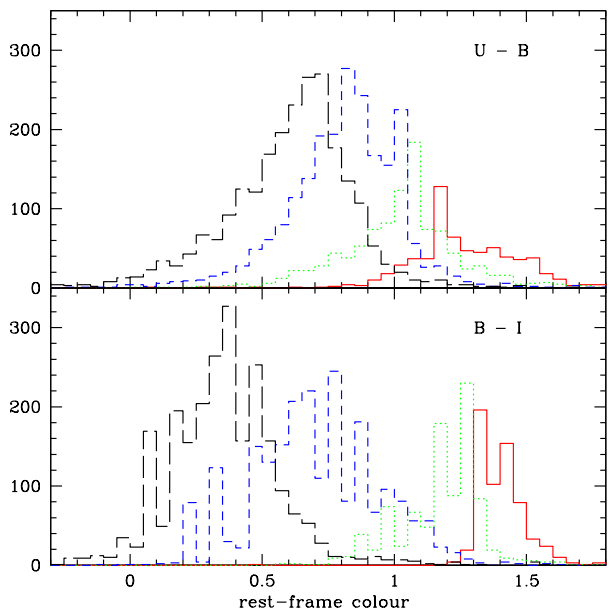}
\caption{
Rest frame $U-B$ (upper panel) and $B-I$ (lower panel) colours for galaxies
of different types: type 1 (solid line), type 2 (dotted line), type 3
(short dashed line), type 4 (long dashed line). 
} 
\label{colors}
\end{figure}

\begin{figure*}
\centering
\includegraphics[width=\hsize]{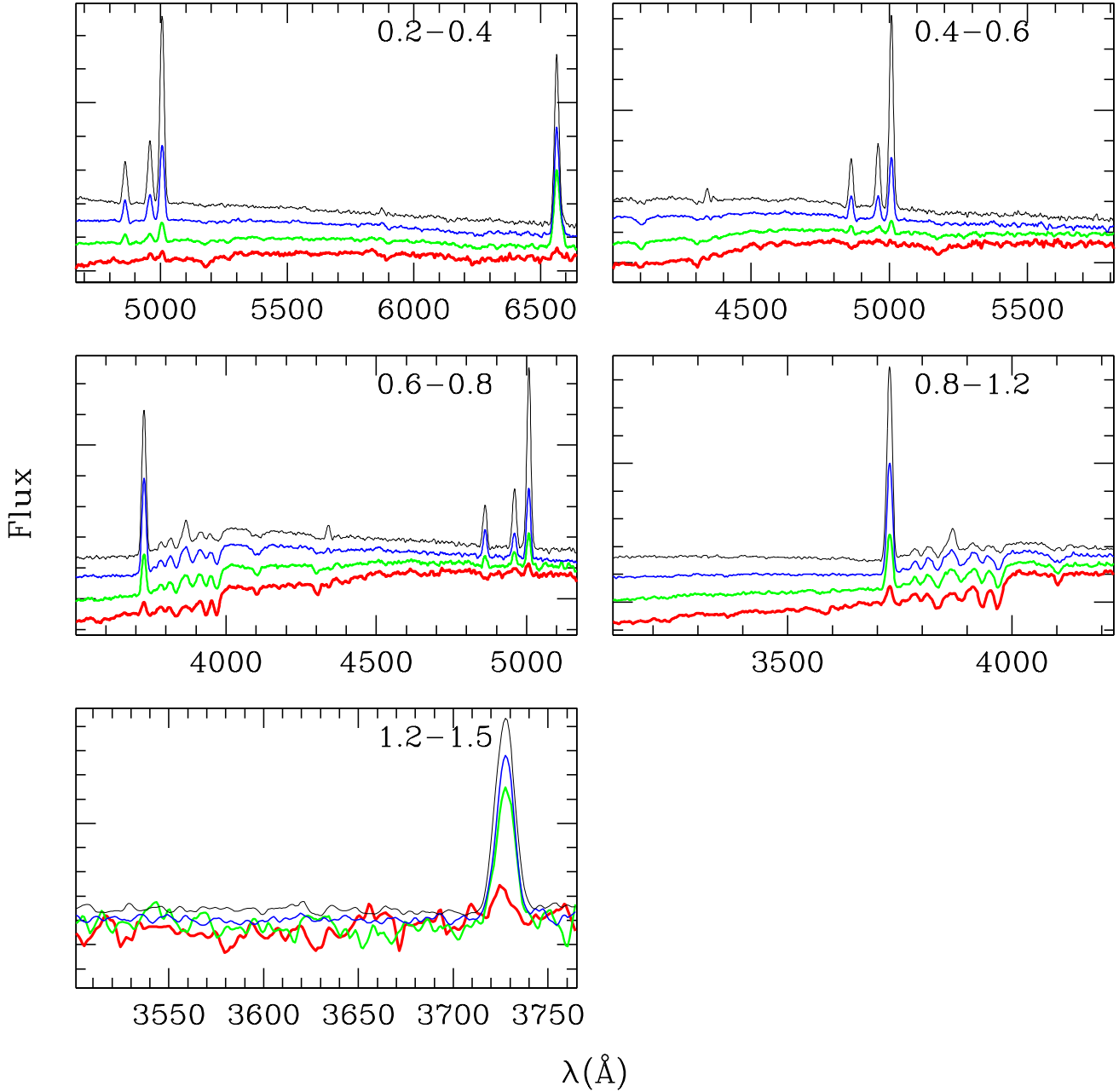}
\caption{
Co-added VVDS spectra of galaxies of the four different types in various
redshift bins, shown in rest frame wavelength. 
The different types are indicated with different line strengths:
the ligther the line, the later the galaxy type.
The redshift bin is indicated in the label of each panel. The flux on the y-axis
is in arbitrary units and the spectra of the various types are arbitrarily 
rescaled for clarity.
} 
\label{class_types}
\end{figure*}
\section{Introduction}

An unbiased and detailed characterization of the luminosity function (LF)
of field galaxies is a basic requirement in many extragalactic issues.
\\
At present the local luminosity function is well constrained by the 
results obtained by the 2dF Galaxy Redshift Survey (2dFGRS, Norberg et al.
\cite{norberg02}) and by the Sloan Digital Sky Survey (SDSS, Blanton et al. 
\cite{blanton03}). These surveys measure redshifts for $10^5 - 10^6$ galaxies 
over a large area, and therefore explore well the properties of the local 
($z<0.3$) Universe. Such large numbers of objects also allow to study
the luminosity functions (as well as the correlation functions and other
properties) for galaxies of different types, defined on the basis of 
colours and/or spectral properties.
A critical analysis of the luminosity functions depending on galaxy type as measured
from the various redshift surveys, as well as a comparison of the different
results, can be found in de~Lapparent (\cite{delapparent03b}).
\\
Madgwick et al. (\cite{madgwick02}), analyzing 2dFGRS data, find a 
systematic steepening of the faint end slope and a faintening of $M^*$
of the luminosity function as one moves from passive to active star 
forming galaxies.
Similar results are found by Blanton et al. (\cite{blanton01}) 
for the SDSS sample, moving from the redder to the bluer galaxies.
\\
For what concerns the high redshift Universe, several studies in the past 
ten years have aimed to map the evolution of the luminosity function.
However, because of the long exposure times required to obtain spectra
of high redshift galaxies, spectroscopic surveys were limited to
a few $10^2$ objects. 
The Canadian Network for Observational Cosmology
field galaxy redshift survey (CNOC-2, Lin et al. \cite{lin99}) and the
ESO Sculptor Survey (ESS, de~Lapparent et al. \cite{delapparent03a}) derived
the luminosity function up to $z\sim 0.5$ using $\sim 2000$ and $\sim 600$
redshifts, respectively. 
de~Lapparent et al. (\cite{delapparent03a}) find a behaviour of the LF
by type similar to the local one derived from 2dFGRS and a strong evolution
of a factor 2 in the volume density of the late type galaxies with respect to
the early type galaxies.
Lin et al. (\cite{lin99}) find for early type galaxies a positive 
luminosity evolution with increasing redshift, which is nearly 
compensated by a negative density evolution. 
On the contrary, for late type galaxies they find a strong
positive density evolution, with nearly no luminosity evolution.
At higher redshift, 
the Canada France Redshift Survey (CFRS, Lilly et al. \cite{lilly95})
allowed to study the luminosity function up to $z\sim 1.1$ with a sample of
$\sim 600$ redshifts. From this survey, the LF of the red population
shows small changes with redshift, while the LF of the blue population 
brightens by about one magnitude from $z\sim 0.5$ to $z\sim 0.75$.
Other results suggest a strong number density evolution of early type
galaxies (Bell et al. \cite{bell04}, Faber et al. \cite{faber05}); 
conversely, in the K20 survey (Cimatti et al. \cite{k20}), Pozzetti
et al. (\cite{pozzetti03}) found that red and early type galaxies
dominate the bright end of the LF and that their number density
shows at most a small decrease ($<30\%$) up to $z\sim 1$ (see also
Saracco et al. \cite{saracco05} and Caputi et al. \cite{caputi06}). 
\\
Luminosity function estimates at higher redshift and/or with larger samples 
are up to now based only on photometric redshifts, like the COMBO-17 survey
(Wolf et al. \cite{wolf03}) and the analysis of the FORS Deep Field (FDF, 
Gabasch et al. \cite{gabasch04}) and the Hubble Deep Fields (HDF-N and HDF-S, 
see e.g. Sawicki et al. \cite{sawicki97}; Poli et al. \cite{poli01}, 
\cite{poli03}); most of these projects derived also the luminosity
function for different galaxy types.
Wolf et al. (\cite{wolf03}) find that early type galaxies show a decrease of 
a factor $\sim 10$ in $\phi^*$ up to $z=1.2$. Latest type galaxies show a 
brightening of about one magnitude in $M^*$ and a $\phi^*$ increase of a 
factor $\sim 1.6$ in their highest redshift bin ($z\sim 1.1$) in the blue band. 
Giallongo et al. (\cite{giallongo05}), using HDFs data, find that the
B band number densities of red and blue galaxies have a different evolution,
with a strong decrease of the red population at $z=2-3$ and a corresponding
increase of the blue population.
Dahlen et al. (\cite{dahlen05}), using GOODS data, claim that the starburst
population fraction increases with redshift by a factor of 3 at $z=2$
in the U band.
\\
Although photometric redshifts represent a powerful tool for deep surveys,
their precision strongly relies on the number of used photometric bands, 
on the templates and on the adopted training procedure; moreover, 
they are affected by the problem
of ``catastrophic errors", i.e. objects with a large difference between
the spectroscopic and the photometric redshift.
\\
A major improvement in this field is obtained with the VIMOS VLT Deep Survey 
(VVDS, Le~F\`evre et al. \cite{lefevre03}) 
and the DEEP-2 Galaxy Redshift Survey (Davis et al. \cite{davis03}). 
The VVDS is an ongoing program to map the evolution of galaxies, 
large scale structures and AGNs from the redshift
measurements of $\sim 10^5$ objects down to a magnitude I$_{AB}=24$, in
combination with a multiwavelength dataset from radio to X-rays.
\\
From the analysis of the evolution of the global luminosity function from the 
first epoch VVDS data (Ilbert et al. \cite{vvdsLF}), 
we found a significant brightening of the
$M^*$ parameter in the U, B, V, R and I rest frame bands, going from
$z=0.05$ to $z=2$. Moreover, we measured an increase of the comoving density
of bright galaxies: this increase depends on the rest frame band, being
higher in the bluest bands.
\\
Among the other results of this survey, we recall the study of the radio
selected objects (Bondi et al. \cite{radio1}) and of their optical
counterparts (Ciliegi et al. \cite{radio2}), the evolution of the
clustering properties (Le~F\`evre et al. \cite{clus1}, Pollo et al.
\cite{clus2}) and of the bias parameter (Marinoni et al. \cite{marinoni05}).
Moreover, from the joined GALEX-VVDS sample, we derived the evolution
of the far UV luminosity function (Arnouts et al. \cite{galexLF}) and
luminosity density (Schiminovich et al. \cite{galexLD}).
\\
In this paper we study the evolution of the luminosity functions 
of galaxies of different spectral types based on the VVDS data. 
This sample allows to perform this analysis for the first time with 
excellent statistical accuracy over a large redshift range ($0.05<z<1.5$).
\\
The plan of the paper is the following: in sect. 2 we briefly present 
the first epoch VVDS 
sample, in sect. 3 we describe the galaxy classification and in sect. 4 we
illustrate the method we used to estimate the luminosity functions. 
In sect. 5 we compare the luminosity functions of the different galaxy types 
and in sect. 6 we show the evolution with redshift of the luminosity functions
by type. Finally in sect. 7 we compare our results
with previous literature estimates and in sect. 8 we summarize our results.
\\
Throughout the paper we adopt the cosmology $\Omega_m = 0.3$ and 
$\Omega_\Lambda = 0.7$, with $h = H_0 / 100$ km s$^{-1}$ Mpc$^{-1}$,
Magnitudes are given in the AB system and are expressed in the five
standard bands U (Bessel), B and V (Johnson), R and I (Cousins). 

\begin{figure}
\centering
\includegraphics[width=\hsize]{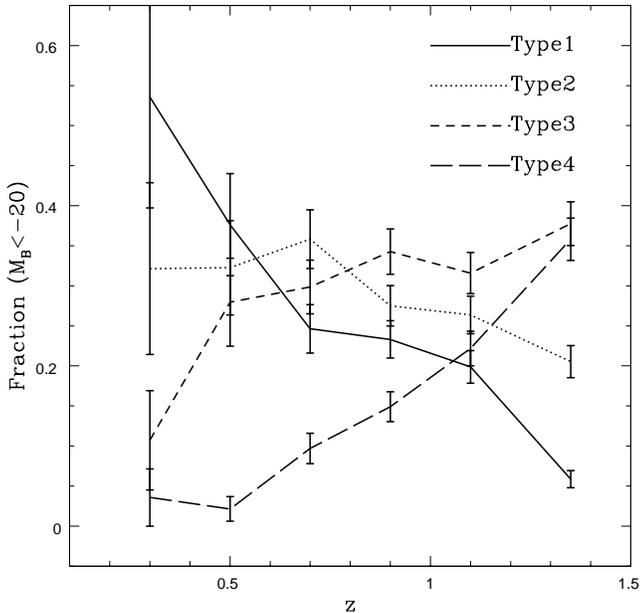}
\caption{Observed fraction of bright galaxies ($M_{B_{AB}}-5log(h) < -20$) 
of different types as a function of redshift. Error bars are $1\sigma$
Poisson errors.
} 
\label{frac_20}
\end{figure}

\section{The first epoch VVDS sample}

The VVDS is described in detail in Le~F\`evre et al. (\cite{vvds1}): here
we report only the main characteristics of the sample used for the 
analysis presented in this paper.
\\
The entire VVDS is formed by a wide part on 4 fields (which is not 
used in this paper), and by a deep part, with spectroscopy in the range 
$17.5\le$ I$_{AB}\le 24$ on the field 0226-04.
Multicolour photometry is available for each field (Le~F\`evre et al. 
\cite{photom1}): in particular, the B, V, R, I photometry for the 0226-04 
deep field is described in detail in McCracken et al. (\cite{photom2}).
Moreover, U band (Radovich et al. \cite{photomU}) and J and K band 
(Iovino et al.
\cite{photomK}) data are available for smaller areas of these fields.
\\ 
Starting from these photometric catalogues, spectroscopic observations
were performed with the VIsible Multi--Object Spectrograph (VIMOS,
Le~F\`evre et al. \cite{vimos}) mounted on the ESO Very Large Telescope
(UT3). The selection of objects for spectroscopic observations 
was based only on magnitude, without any other colour or shape criteria.
\\
Deep spectroscopic observations ($17.5\le$ I$_{AB}\le 24$) were performed 
also on the Chandra Deep Field South (VVDS-CDFS, Le~F\`evre et al. 
\cite{vvds-cdfs}), starting from the EIS I band photometry and astrometry 
(Arnouts et al. \cite{eis}). Multicolour U, B, V, R and I photometry 
for this sample is available from the COMBO-17 survey 
(Wolf et al. \cite{wolf03}).
\\
Spectroscopic data were reduced with the VIMOS Interactive Pipeline 
Graphical Interface (VIPGI, Scodeggio et al. \cite{vipgi}, Zanichelli et al.
\cite{ifu}) and redshift measurements were performed with the KBRED package
(Scaramella et al. \cite{kbred}) and then visually checked. Each redshift
measurement was assigned a quality flag, ranging from 0 (failed measurement)
to 4 (100\% confidence level); flag 9 indicates spectra with a single 
emission line, for which multiple solutions are possible. Further details
on the quality flags are given in Le~F\`evre et al. (\cite{vvds1}).
\\
The analysis presented in this paper is based on the first epoch VVDS deep
sample, which has been obtained from the first observations (fall 2002) on
the fields VVDS-02h and VVDS-CDFS, which cover 1750 and 450 arcmin$^2$,
respectively.
We eliminated from the sample spectroscopically confirmed stars and
broad line AGNs, remaining with 6477 + 1236 galaxy spectra with secure 
spectroscopic identification (flag 2, 3, 4, 9), corresponding to a confidence 
level higher than 75\%.
Redshifts with flags 0 and 1 are taken into account statistically (see
sect. 4). This spectroscopic sample, which is purely magnitude selected,
has a median redshift of $\sim 0.76$.

\section{Galaxy classification}

Galaxies have been classified using all the multicolour information available;
in the VVDS-02h field B, V, R and I band magnitudes are available for
all galaxies, while U band data are available for 83\% of the galaxies.
For the VVDS-CDFS sample U, B, V, R and I photometry from the
COMBO-17 survey is used.
\\
Absolute magnitudes are computed following the method described in
the Appendix of Ilbert et al. (\cite{vvdsLF}). The K-correction is computed
using a set of templates and all the photometric information (UBVRI)
available. However, in order to reduce the template dependency, the
rest frame absolute magnitude in each band is derived using the apparent 
magnitude from the closest observed band, redshifted at the redshift of the
galaxy. With this method, the applied K-correction is as small as possible 
as possible. 
\\
For each galaxy the rest frame magnitudes were matched with the empirical
set of SEDs described in Arnouts et al. (\cite{arnouts99}), composed of
four observed spectra (CWW, Coleman et al. \cite{cww}) and two starburst
SEDs computed with GISSEL (Bruzual \& Charlot \cite{bc93}).
The match is performed minimizing a $\chi^2$ variable on these templates
at the spectroscopic redshift of each galaxy.
The same procedure has been applied by Lin et al. (\cite{lin99}) to the
CNOC-2 survey up to $z\sim 0.55$.
This approach is also similar to that adopted by Wolf et al. (\cite{wolf03})
for the COMBO-17 survey, but we have the advantage of using spectroscopic
redshifts, while they had to rely on photometric redshifts.
\\
Galaxies have been divided in four types, corresponding to the E/S0 template 
(type 1), early spiral template (type 2), late spiral template (type 3) and 
irregular template (type 4). 
These types are based on the four CWW templates: type 4 includes also the two
starburst templates. 
The numbers of galaxies for each type are listed in Table \ref{numbers}.
\\
In order to have an idea of the correspondence of these types with colours,
we report here the rest frame colours for each template: 
type 1, 2, 3 and 4 have $B_{AB}-I_{AB}=$1.58, 1.11, 0.79 and 0.57, 
respectively.
Given these colours, a rough colour subdivision for each class is
$1.3<B_{AB}-I_{AB}$ for type 1, $0.95<B_{AB}-I_{AB}<1.3$ for
type 2, $0.68<B_{AB}-I_{AB}<0.95$ for type 3 and
$B_{AB}-I_{AB}<0.68$ for type 4.
However, we remind that these colour ranges are only indicative and
our classification scheme is based on the whole multicolour 
coverage.
In Fig.\ref{colors} we show the $U-B$ and $B-I$ colour distributions for 
the galaxies of our sample divided according to type. From this figure it is clear
that, although the different types have different colour distributions, they
present significant overlaps. This fact is a consequence of classification 
schemes using template fitting on multicolour data.
\\
Note that, in order to avoid to be model dependent, we did not apply 
to the templates any correction aimed at taking into account colour 
evolution with redshift. 
It is well known that the colour of a simple stellar population subject
to passive evolution was bluer in the past. In principle, this could 
imply that galaxies classified as type 1 at low redshift might be
classified differently at higher redshift. 
Indeed, this effect has been invoked by the authors who found negative
evolution in the luminosity function of ``red" galaxies 
(see f.i. Wolf et al. \cite{wolf03}). 
In order to verify this hypothesis, we applied our classification scheme
to synthetic spectra (Bruzual \& Charlot \cite{bc93}) of ellipticals 
(i.e. simple stellar populations and exponentially declining star formation 
with time scales of 0.1 Gyr and 0.3 Gyr) with formation redshift between
$z_{form}=$2 and 20. We find that all ellipticals with $z_{form}> 2$  
would be classified as type 1 objects, even at $z\sim 1$.   
\\
In order to check, at least on a statistical basis, the consistency
between this photometric classification and average spectral properties,
we summed the
normalized spectra of all galaxies in each of the four types. 
The resulting average spectra are shown in Fig.\ref{class_types} for each
type in various redshift bins. This figure confirms the robustness of our
classification scheme: moving from type 1 to type 4 objects, 
the composite spectra show an increasingly blue continuum, with emission 
lines of increasing strength.
This confirms that the four types show different spectral features and
therefore represent different classes of objects.
\\
For the VVDS-CDFS sample, HST-ACS images are available. Using these data,
Lauger et al. (\cite{lauger06}) classified the galaxies in this sample
using an asymmetry-concentration diagram. 
Plotting our type 1 galaxies in this diagram, we find that $\sim 91\%$ of
them lie in the region of bulge dominated objects, showing an excellent
consistency also between our photometric classification and a 
morphological one. 
\\
In Fig.\ref{frac_20} we plot the observed fraction of bright galaxies 
of each type as a function of redshift. We selected objects with 
$M_{B_{AB}}-5log(h) < -20$ because these galaxies are visible in the whole 
redshift range. From this figure it is clear the growing importance of 
bright late type objects with increasing redshift and the corresponding 
strong decrease of the fraction of bright early type galaxies.

\begin{table}
\caption[]{Numbers of galaxies of different types in the sample}
\begin{flushleft}
\begin{tabular}{rr}
\hline\noalign{\smallskip}
       Type      &  N$_{gal}$  \\
      \noalign{\smallskip}
      \hline
      \noalign{\smallskip}
       total        & 7713   \\
       type 1       & ~730   \\
       type 2       & 1290   \\
       type 3       & 2622   \\
       type 4       & 3071   \\
\noalign{\smallskip}
\hline
\end{tabular}
\end{flushleft}
\label{numbers}
\end{table}

\begin{figure*}
\centering
\includegraphics[width=0.45\hsize]{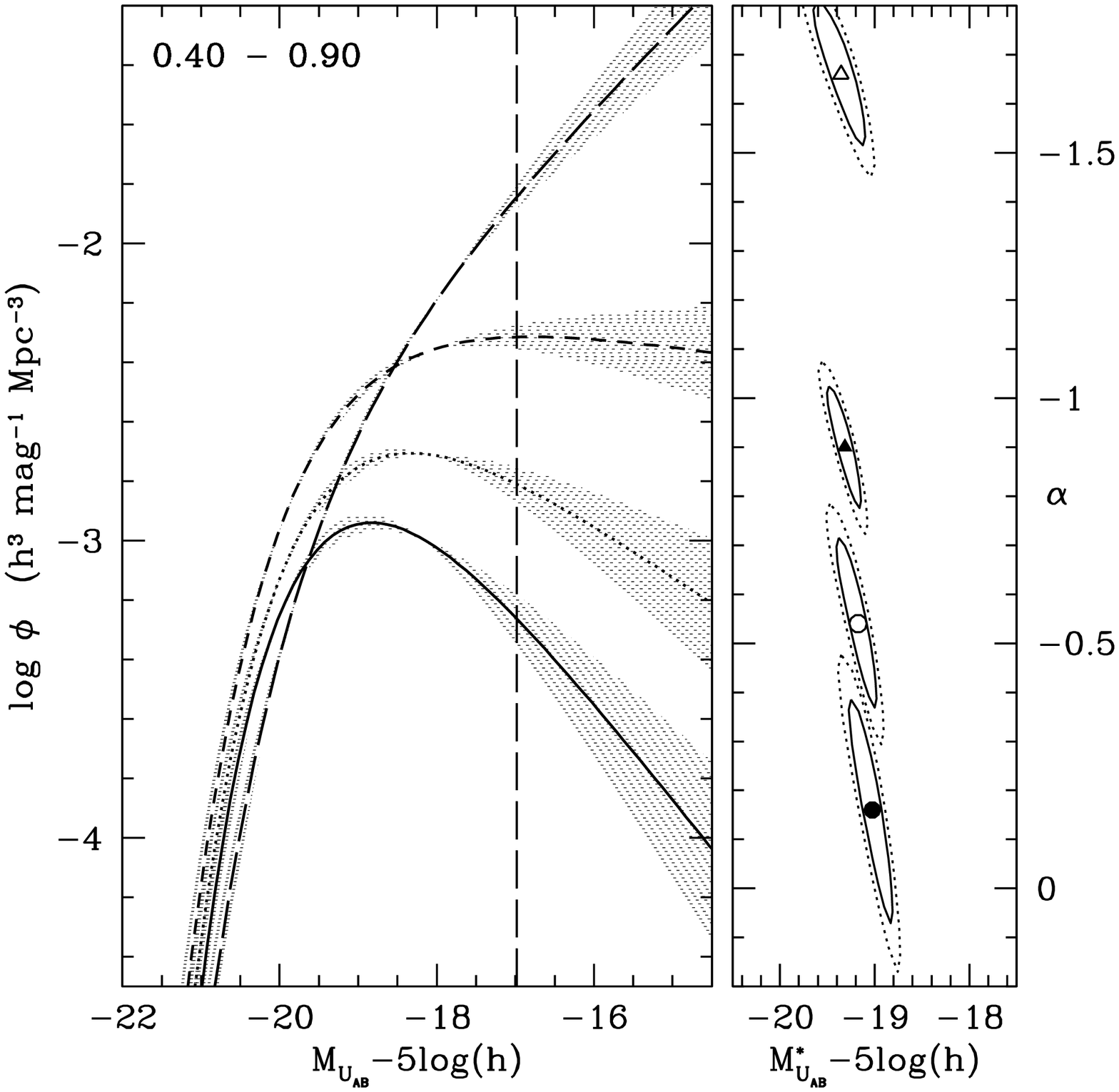}
\includegraphics[width=0.45\hsize]{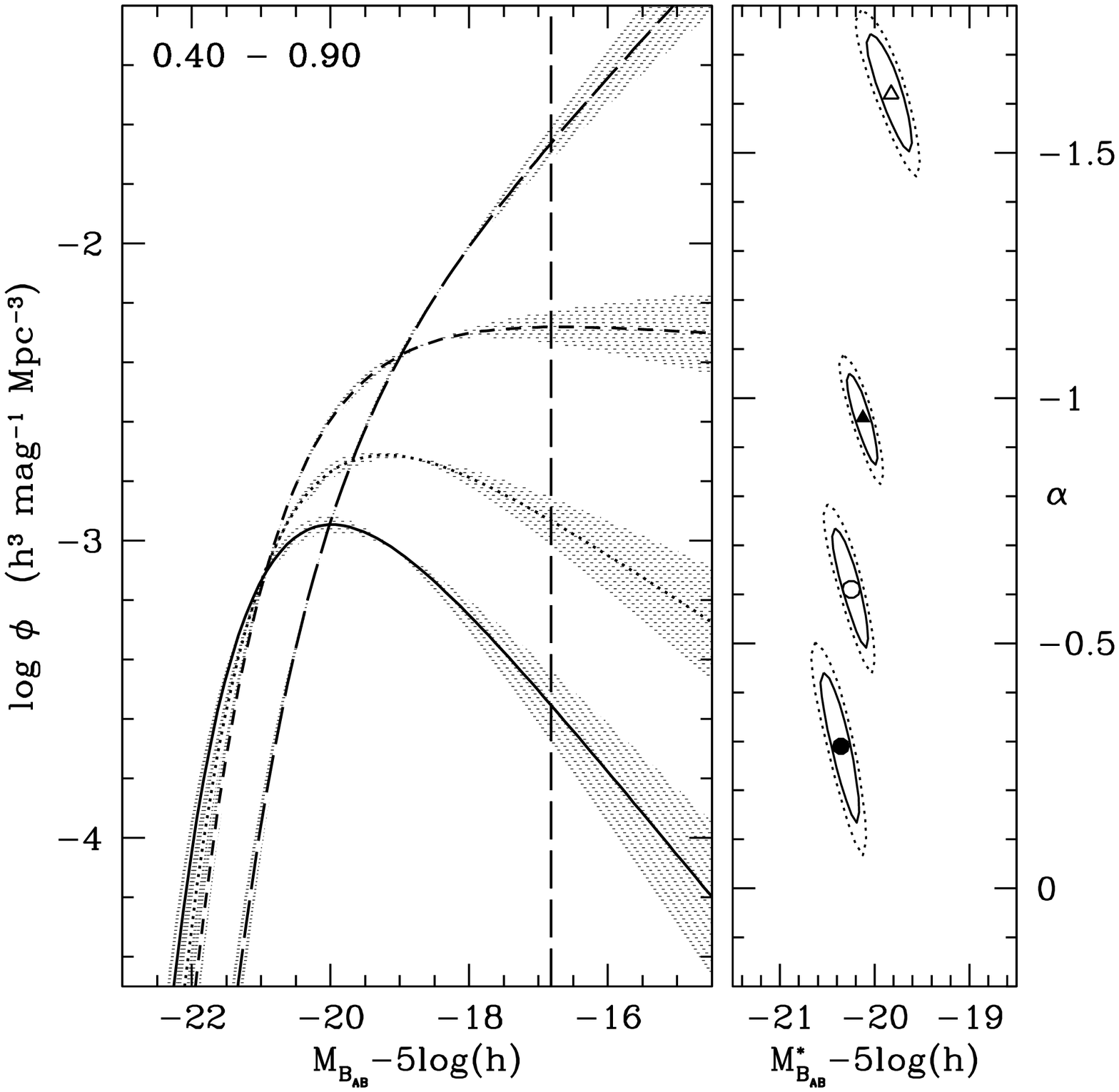}
\includegraphics[width=0.45\hsize]{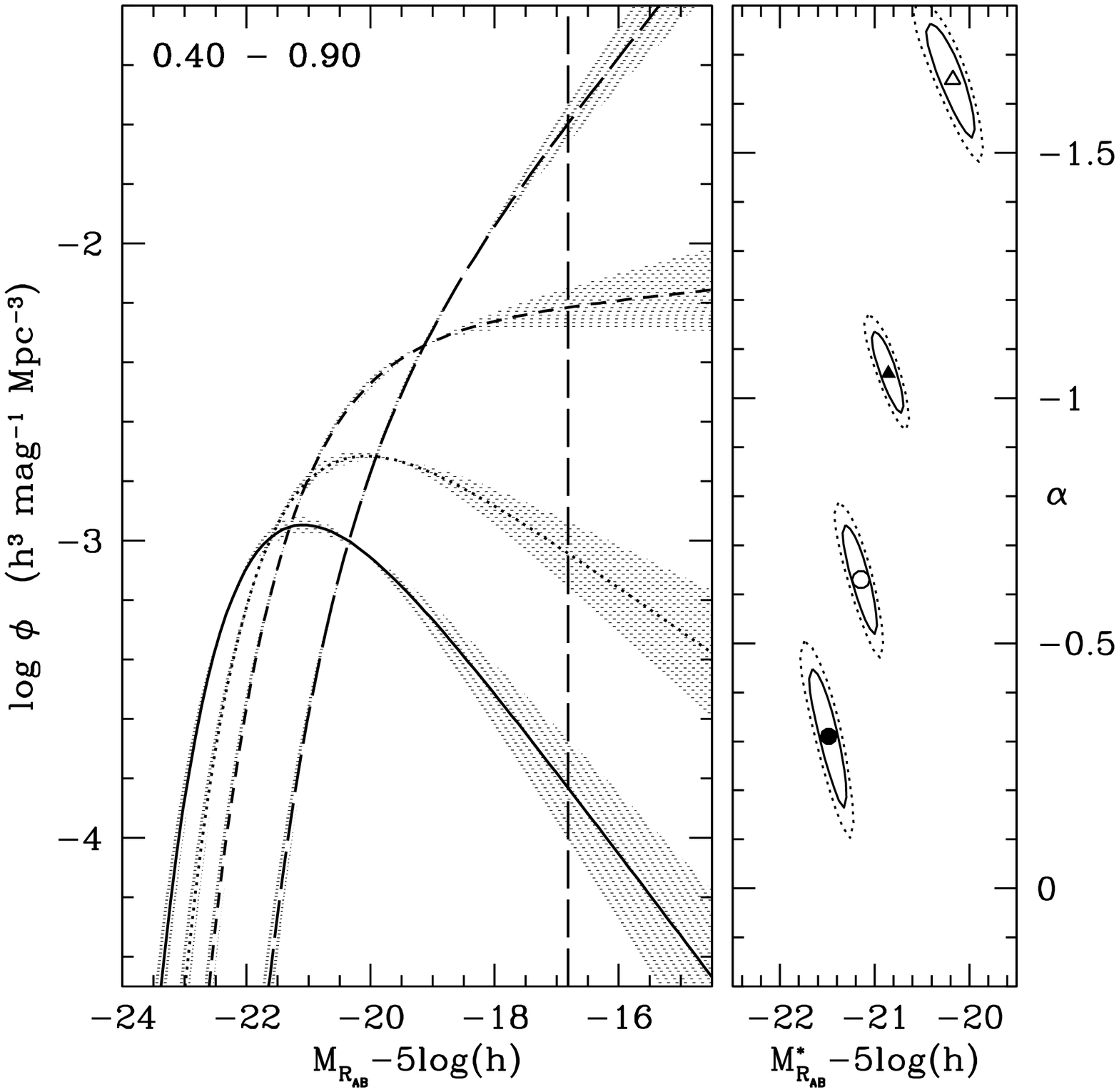}
\includegraphics[width=0.45\hsize]{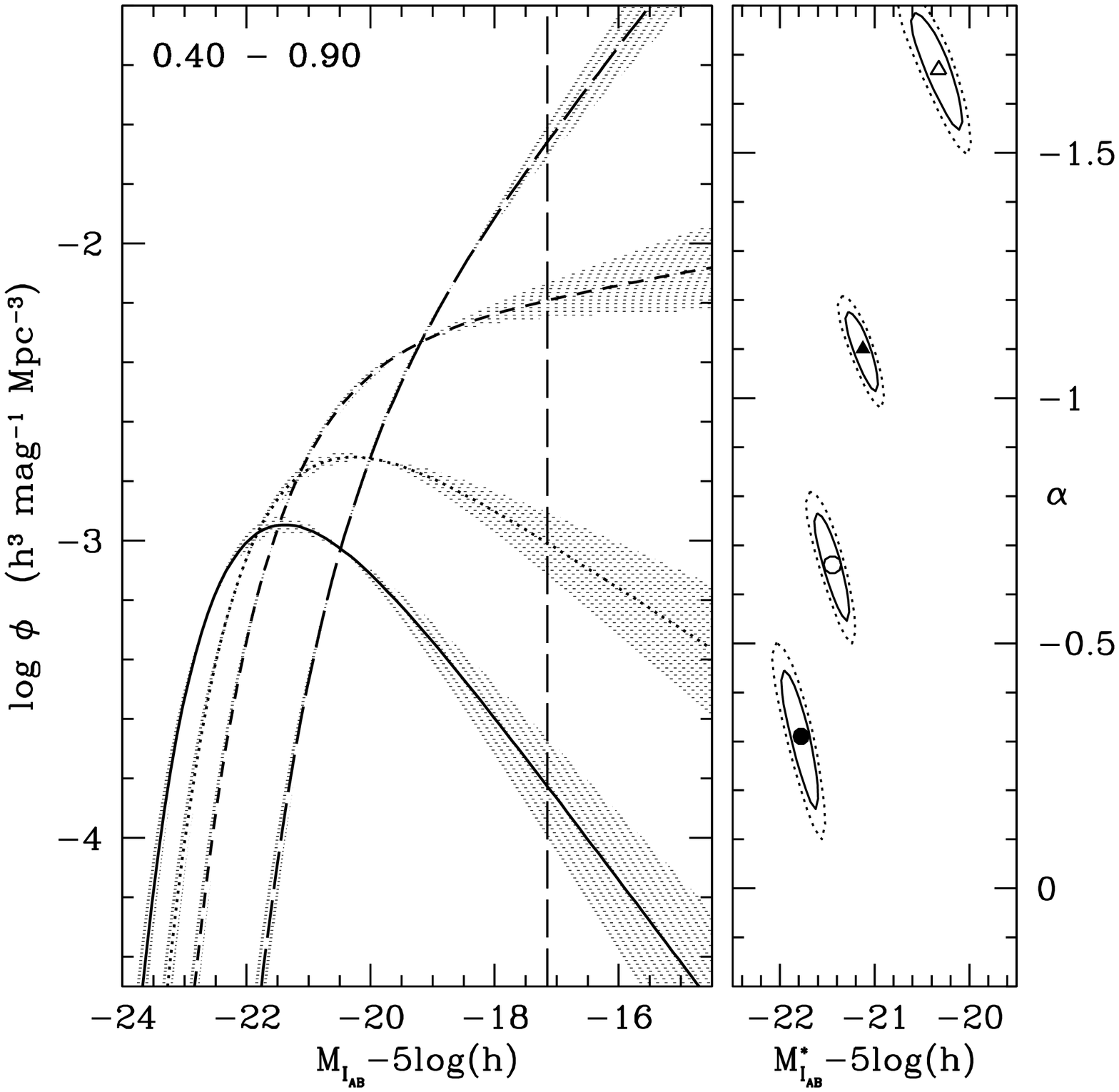}
\caption{
Luminosity functions by different types in the redshift range
$[0.4-0.9]$ in various rest frame bands: upper panels U (left) and B (right);
lower panels R (left) and I (right). Note that the ranges of magnitude are 
different in the various panels. The lines represent the
$STY$ estimates for type 1 (solid), type 2 (dotted), type 3 (short dashed)
and type 4 (long dashed) galaxies. 
The vertical dashed line represents the faint absolute magnitude limit 
considered in the $STY$ estimate (see text). The shaded regions represent the 68\%
uncertainties on the parameters $\alpha$ and $M^*$, whose confidence ellipses 
are reported in the right panels. The ellipse contours are at 68\% and 90\%
confidence level (solid and dotted line respectively). The points inside the
ellipses represent
the best fit values for type 1 (filled circle), type 2 (open circle), type 3
(filled triangle) and type 4 (open triangle) galaxies.     
}
\label{LF_04-09}
\end{figure*}

\begin{table*}
\caption[]{$STY$ parameters (with $1\sigma$ errors) for different galaxy types 
in different bands in the redshift range $[0.4 - 0.9]$ }
\begin{flushleft}
\begin{tabular}{c c c c c c c} \hline

\multicolumn{7}{c}{$\Omega_m$=0.3 \hspace{1cm}               $\Omega_\Lambda$=0.7} \\ \hline
Band & Type &  Number$^{(a)}$ &  Number$^{(b)}$ &   $\alpha$ &  $M^*_{AB}-5log(h)$ & $\phi^*$($10^{-3} h^3 Mpc^{-3}$)   \vspace{0.2cm} \\ \hline
\hline

U  &    1    &   411 &   357 &  -0.16$^{{\rm + 0.15}}_{{\rm - 0.15}}$ & -19.03$^{{\rm + 0.14}}_{{\rm - 0.15}}$ &   3.35$^{{\rm + 0.20}}_{{\rm - 0.26}}$ \\ 
   &    2    &   677 &   596 &  -0.54$^{{\rm + 0.12}}_{{\rm - 0.11}}$ & -19.18$^{{\rm + 0.13}}_{{\rm - 0.14}}$ &   4.83$^{{\rm + 0.48}}_{{\rm - 0.53}}$ \\ 
   &    3    &  1371 &  1192 &  -0.90$^{{\rm + 0.08}}_{{\rm - 0.08}}$ & -19.31$^{{\rm + 0.11}}_{{\rm - 0.12}}$ &   7.32$^{{\rm + 0.86}}_{{\rm - 0.86}}$ \\ 
   &    4    &  1442 &  1238 &  -1.66$^{{\rm + 0.10}}_{{\rm - 0.10}}$ & -19.35$^{{\rm + 0.17}}_{{\rm - 0.18}}$ &   4.09$^{{\rm + 1.21}}_{{\rm - 1.04}}$ \\ 

B  &    1    &   411 &   404 &  -0.29$^{{\rm + 0.10}}_{{\rm - 0.10}}$ & -20.35$^{{\rm + 0.13}}_{{\rm - 0.13}}$ &   3.19$^{{\rm + 0.23}}_{{\rm - 0.26}}$ \\ 
   &    2    &   677 &   669 &  -0.61$^{{\rm + 0.08}}_{{\rm - 0.08}}$ & -20.25$^{{\rm + 0.12}}_{{\rm - 0.12}}$ &   4.48$^{{\rm + 0.43}}_{{\rm - 0.44}}$ \\ 
   &    3    &  1371 &  1349 &  -0.96$^{{\rm + 0.06}}_{{\rm - 0.06}}$ & -20.12$^{{\rm + 0.10}}_{{\rm - 0.11}}$ &   6.79$^{{\rm + 0.72}}_{{\rm - 0.71}}$ \\ 
   &    4    &  1442 &  1403 &  -1.62$^{{\rm + 0.08}}_{{\rm - 0.08}}$ & -19.83$^{{\rm + 0.15}}_{{\rm - 0.16}}$ &   4.46$^{{\rm + 1.08}}_{{\rm - 0.95}}$ \\ 

V  &    1    &   411 &   411 &  -0.31$^{{\rm + 0.09}}_{{\rm - 0.09}}$ & -21.13$^{{\rm + 0.12}}_{{\rm - 0.13}}$ &   3.16$^{{\rm + 0.23}}_{{\rm - 0.25}}$ \\ 
   &    2    &   677 &   677 &  -0.61$^{{\rm + 0.07}}_{{\rm - 0.07}}$ & -20.82$^{{\rm + 0.11}}_{{\rm - 0.12}}$ &   4.51$^{{\rm + 0.40}}_{{\rm - 0.41}}$ \\ 
   &    3    &  1371 &  1371 &  -1.00$^{{\rm + 0.06}}_{{\rm - 0.06}}$ & -20.57$^{{\rm + 0.10}}_{{\rm - 0.11}}$ &   6.21$^{{\rm + 0.66}}_{{\rm - 0.64}}$ \\ 
   &    4    &  1442 &  1442 &  -1.62$^{{\rm + 0.08}}_{{\rm - 0.08}}$ & -20.03$^{{\rm + 0.15}}_{{\rm - 0.16}}$ &   4.36$^{{\rm + 1.03}}_{{\rm - 0.92}}$ \\ 

R  &    1    &   411 &   411 &  -0.31$^{{\rm + 0.09}}_{{\rm - 0.09}}$ & -21.48$^{{\rm + 0.12}}_{{\rm - 0.13}}$ &   3.16$^{{\rm + 0.23}}_{{\rm - 0.25}}$ \\ 
   &    2    &   677 &   677 &  -0.63$^{{\rm + 0.07}}_{{\rm - 0.07}}$ & -21.15$^{{\rm + 0.11}}_{{\rm - 0.12}}$ &   4.37$^{{\rm + 0.40}}_{{\rm - 0.41}}$ \\ 
   &    3    &  1371 &  1371 &  -1.05$^{{\rm + 0.05}}_{{\rm - 0.05}}$ & -20.86$^{{\rm + 0.11}}_{{\rm - 0.11}}$ &   5.55$^{{\rm + 0.63}}_{{\rm - 0.60}}$ \\ 
   &    4    &  1442 &  1436 &  -1.65$^{{\rm + 0.08}}_{{\rm - 0.08}}$ & -20.18$^{{\rm + 0.16}}_{{\rm - 0.17}}$ &   3.87$^{{\rm + 0.99}}_{{\rm - 0.87}}$ \\ 

I  &    1    &   411 &   411 &  -0.31$^{{\rm + 0.09}}_{{\rm - 0.09}}$ & -21.78$^{{\rm + 0.12}}_{{\rm - 0.13}}$ &   3.17$^{{\rm + 0.23}}_{{\rm - 0.25}}$ \\ 
   &    2    &   677 &   677 &  -0.66$^{{\rm + 0.07}}_{{\rm - 0.07}}$ & -21.44$^{{\rm + 0.12}}_{{\rm - 0.12}}$ &   4.22$^{{\rm + 0.40}}_{{\rm - 0.40}}$ \\ 
   &    3    &  1371 &  1370 &  -1.10$^{{\rm + 0.05}}_{{\rm - 0.05}}$ & -21.13$^{{\rm + 0.11}}_{{\rm - 0.11}}$ &   5.05$^{{\rm + 0.60}}_{{\rm - 0.57}}$ \\ 
   &    4    &  1442 &  1413 &  -1.67$^{{\rm + 0.08}}_{{\rm - 0.08}}$ & -20.32$^{{\rm + 0.17}}_{{\rm - 0.18}}$ &   3.58$^{{\rm + 0.97}}_{{\rm - 0.86}}$ \\ 

\hline
\multicolumn{7}{l}{(a) Number of galaxies in the redshift bin }\\
\multicolumn{7}{l}{(b) Number of galaxies brighter than the bias limit 
(sample used for $STY$ estimate; see the text for details)}\\

\end{tabular}
\end{flushleft}
\label{param04-09}
\end{table*}

\section{Luminosity function estimate}
 
Luminosity functions were derived using the Algorithm for Luminosity Function 
(ALF), a dedicated tool which uses various estimators: the non-parametric 
$1/V_{max}$ (Schmidt \cite{schmidt68}), $C^+$ (Lynden Bell \cite{lyndenbell71}),
$SWML$ (Efstathiou et al. \cite{swml}) and the parametric $STY$ (Sandage,
Tammann \& Yahil \cite{sty}), for which we assumed
a Schechter function (Schechter \cite{schechter76}).
The tool and these estimators, as well as their specific use in the context 
of the VVDS, are described in detail in Ilbert et al. (\cite{vvdsLF}).
\\
Ilbert et al. (\cite{ilbert04}) have shown that the estimate of the global
luminosity function can be biased, mainly in its faint end, when the
band in which it is measured is far from the rest frame band in which
galaxies are selected. This is due to the fact that, because of the
K-corrections, different galaxy types are visible in different absolute
magnitude ranges at a given redshift. When computing the global luminosity
functions (Ilbert et al. \cite{vvdsLF}), we avoided this bias by using
for the $STY$ estimate, in each redshift range, only galaxies within the 
absolute magnitude range where all the SEDs are potentially observable.
\\
Even if this bias is much less important when estimating the luminosity 
function of galaxies divided by types, we have, however, taken it into 
account.
The absolute magnitude limits for the $STY$ estimate are indicated with
vertical dashed lines in the figures, and in the tables where the best
fit parameters are reported (Table \ref{param04-09} and \ref{parameters})
we give both the total number of objects and the number of galaxies within
this magnitude limit. 
\\
In order to take into account the unknown redshifts (not 
observed objects and failed spectra), a weight was applied to each galaxy, 
following the procedure described in detail in Ilbert et al. (\cite{vvdsLF}).
\\
This weight is a combination of two different contributions: the target 
sampling rate and the spectroscopic success rate.
The target sampling rate, i.e. the fraction of observed galaxies, 
corrects for the selection effects due to
the procedure used for the mask preparation (Bottini et al. \cite{vmmps}):
to maximize the number of slits, the procedure tends to select objects
with smaller angular size on the x-axis of the image, corresponding to the 
direction in which the slits are placed. As a consequence, the final
spectroscopic sample has a bias against large objects, which produces a 
mild dependence of the target sampling rate on the apparent magnitude.
The target sampling rate is $\sim 25\%$ for most of the sample and
is computed as a function of the object size (see Ilbert et al.
\cite{vvdsLF} for further details).
The spectroscopic success rate takes into account the fraction of
objects without a good redshift determination (i.e. flags 0 and 1).
As shown in Ilbert et al. (\cite{vvdsLF}), these objects are expected to
have a different redshift distribution with respect to that of the
sample with measured redshift, as confirmed by the use of their 
photometric redshifts.
Given the fact that the spectroscopic success rate decreases for 
faint apparent magnitudes, we derived it in four magnitude bins as a
function of redshift (using photometric redshifts, see Fig.3 in 
Ilbert et al. \cite{vvdsLF}). The
shape of the spectroscopic success rate is similar in all magnitude
bins, showing a maximum at $z\sim 0.7$ and two minima for
$z<0.5$ and $z>1.5$. Since the number of galaxies for each type is not
large enough to reliably estimate the spectroscopic success rate as a 
function also of the galaxy type, we have used the global spectroscopic
success rate for all galaxy types.
\\
In order to check the effect of the ``cosmic variance", i.e. variations
in the luminosity function due to fluctuations in the large scale structure,
we applied the following test on the VVDS-02h deep area. 
For this field we derived photometric redshifts (Ilbert et al. \cite{zphot}) 
based on both VVDS photometry (BVRIJK) and on new CFHT Legacy Survey 
photometry (ugriz), which has now become available in a field 
covering 1 sq. deg. (http://www.cfht.hawaii.edu/Science/CFHLS/) which includes 
the 1700 arcmin$^2$ area covered by the VVDS spectroscopic survey.
Then we divided the field in two non overlapping regions (the sub-area where 
spectroscopic data are available and the remaining area) 
and we compared the luminosity distributions of galaxies in these two
samples ($\sim$0.5 sq.deg. each), in the same redshift bins in which the 
luminosity functions were derived.
In each redshift bin the two distributions show average differences of the 
order of 10\%, with some larger fluctuations due to Poisson statistics, 
without any systematic trend. Therefore the influence of 
the ``cosmic variance" is expected to be limited. 

\section{Comparison of the luminosity functions of different types}

As a first step, we compare the luminosity functions for galaxies 
of different types, in order to see which is the relative behaviour of the 
various populations.
To perform this comparison, we selected galaxies in the redshift 
range $[0.4 - 0.9]$. About $50\%$ of the objects of our sample are included
in this redshift interval, covering a wide
range of luminosities (i.e. absolute magnitudes in the B band are in the
range $[-23.7; -16.8] - 5$ log$(h)$).
Moreover, the spectroscopic success rate of our survey reaches a maximum at
$z\sim 0.7$ and therefore the possible dependency of the estimated LF 
on the weighting scheme described above is minimized in this redshift interval. 
\\
In Fig.\ref{LF_04-09} we report the luminosity functions estimated with 
the $STY$ method in the rest frame bands U, B, R and I, with the
corresponding confidence ellipses for the $\alpha$ and $M^*$ parameters.
The values of the parameters with their $1\sigma$ errors, as well as the number
of galaxies for each type, are reported in Table \ref{param04-09}.
The $\phi^*$ parameters listed in this table are derived adopting 
the density estimator of Efstathiou et al. (\cite{swml}), following the
procedure described in the Appendix of Ilbert et al. (\cite{vvdsLF}). 
\\
Table \ref{param04-09} shows that our estimates are based on several hundreds 
of galaxies for each type, and are therefore well statistically constrained,
as can be seen also from the sizes of the confidence ellipses in the figures.  
The first, very clear result which appears from Fig.\ref{LF_04-09}
is the significant strong steepening of the luminosity functions going from 
early to late types. In all bands the power law slope steepens by $\Delta
\alpha \sim 1.3 - 1.5$ going from type 1 to type 4 galaxies and
galaxies of late types are the dominant population at faint magnitudes.
\\
Systematic trends are also seen in the $M^*$ parameter. In the reddest bands
(lower panels in Fig.\ref{LF_04-09}), $M^*$ is significantly fainter for late 
type galaxies and this faintening is particularly apparent for type 4 objects.
The brighter $M^*$ for early type galaxies reflects the fact 
that most of the more massive objects belong to this population.
\\
The difference of the $M^*$ values for different types decreases in the 
B band and disappears or even changes sign in the U band. 
This behaviour is explained by the fact that the luminosity in the bluer
bands is dominated by the light of young stars, produced during the star 
formation activity. Galaxies of later types, which are still actively forming 
stars, are therefore more luminous in the bluer bands.
\\
These results are qualitatively in agreement with previous results from
the literature, most of which at lower redshift (see de~Lapparent 
\cite{delapparent03b} for a review of the results from a number of
surveys in the redshift range $0\le z \le 0.6$).
In particular, in almost all surveys the luminosity function of late type
galaxies is steeper and with a fainter $M^*$ with respect to that of early
type galaxies. 
However, a quantitative comparison with previous results is difficult, 
because of the different classification schemes adopted in the various surveys, 
the different redshift ranges and selection criteria.

\begin{figure*}
\centering
\includegraphics[width=\hsize]{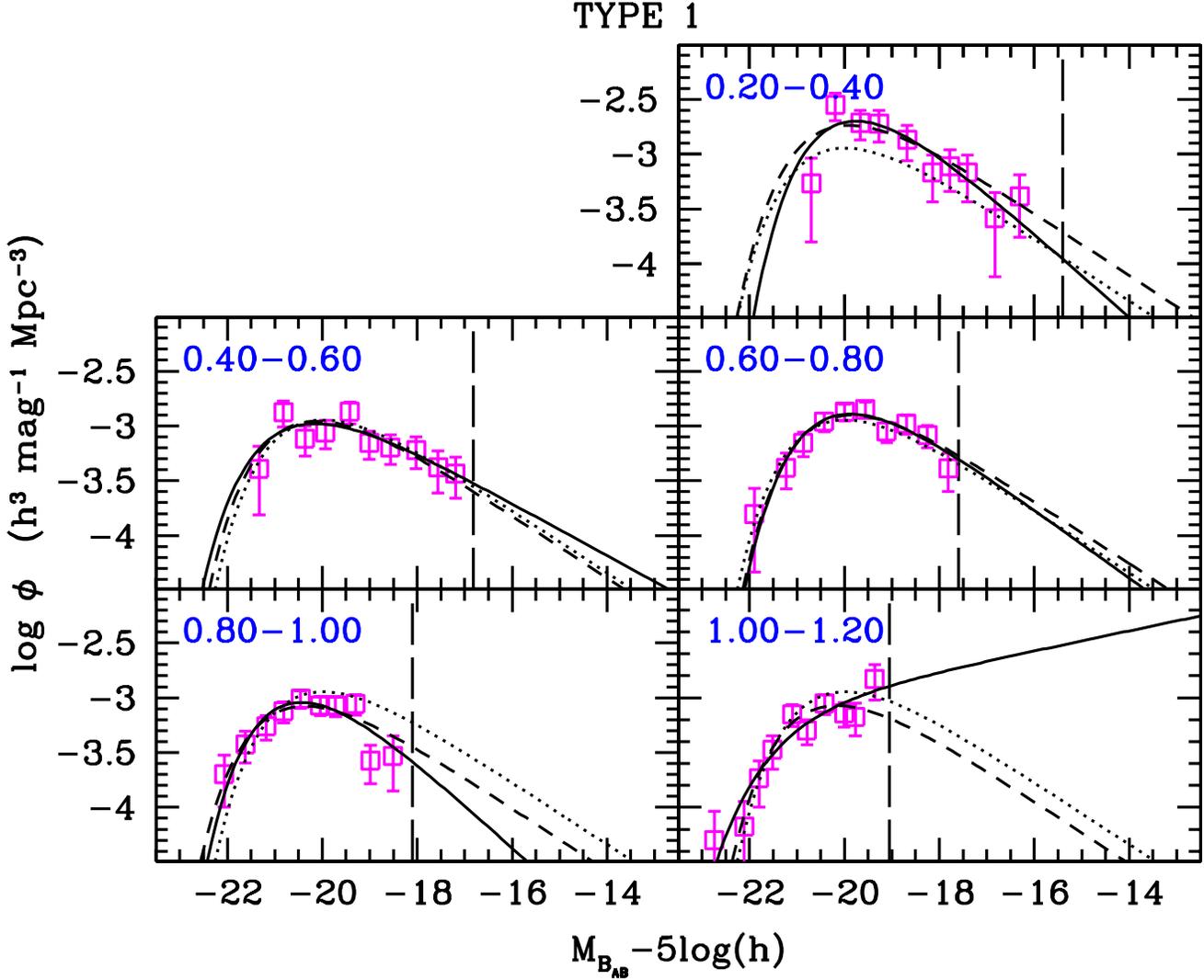}
\caption{
Evolution of the luminosity function in the B-band for type 1 galaxies.
Each panel refers to a different redshift bin, which is indicated in the
label. 
The vertical dashed line represents the faint absolute limit considered in
the $STY$ estimate.
The luminosity functions are estimated with different methods (see text for
details) but for clarity we plot only the results from $C^+$ (open squares), 
and $STY$ (solid line).
The dashed line is the $STY$ estimate obtained by fixing $\alpha$ to the value
determined in the redshift range $[0.4-0.9]$.
The dotted line represents the luminosity function estimated in the
redshift range $[0.4-0.9]$: this curve is reported in each panel as a
reference.
} 
\label{LFtype1}
\end{figure*}

\begin{figure*}
\centering
\includegraphics[width=\hsize]{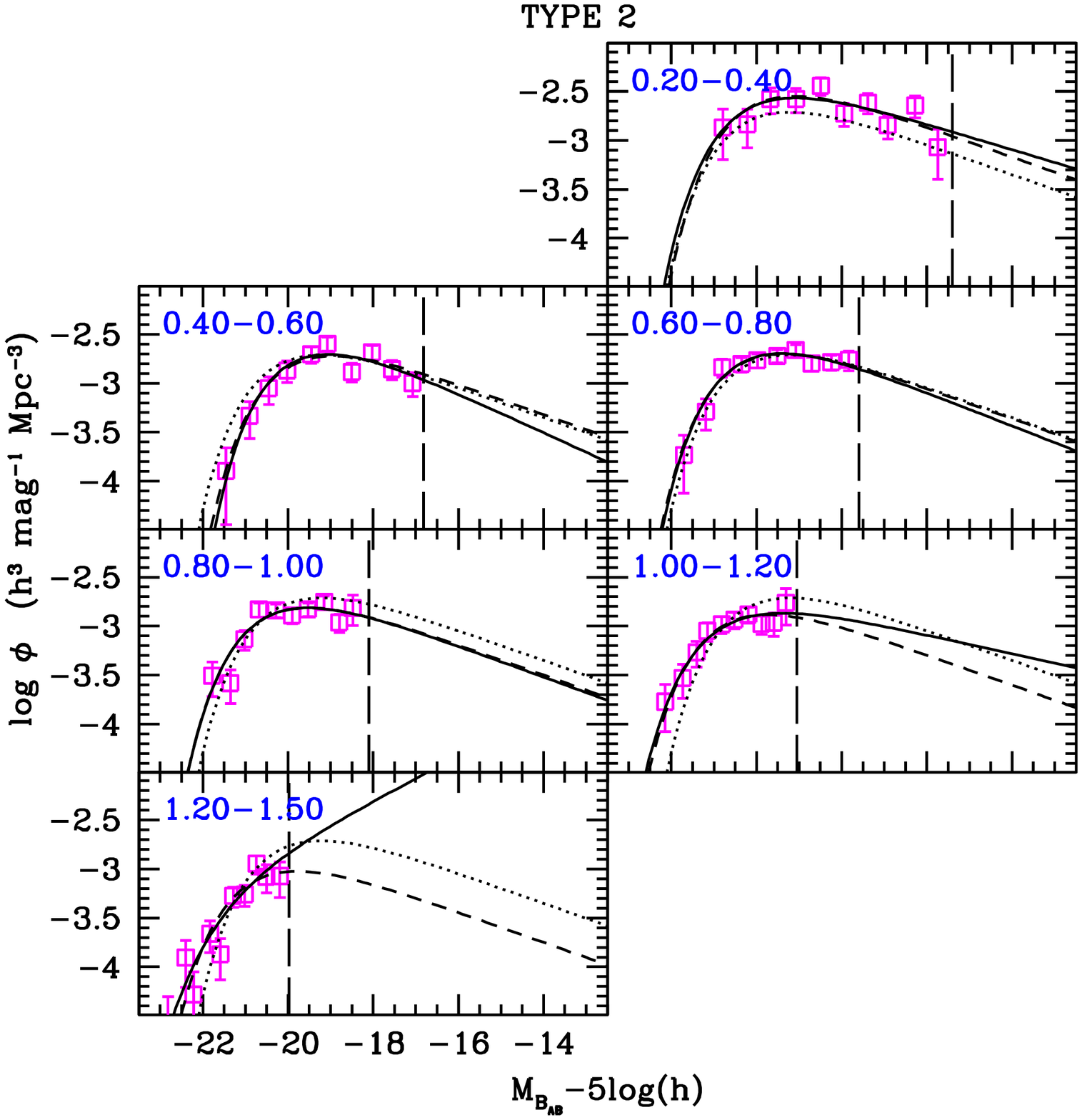}
\caption{
Evolution of the luminosity function in the B-band for type 2 galaxies.
The meaning of the lines and the symbols is the same as in Fig.\ref{LFtype1}.
} 
\label{LFtype2}
\end{figure*}

\begin{figure*}
\centering
\includegraphics[width=\hsize]{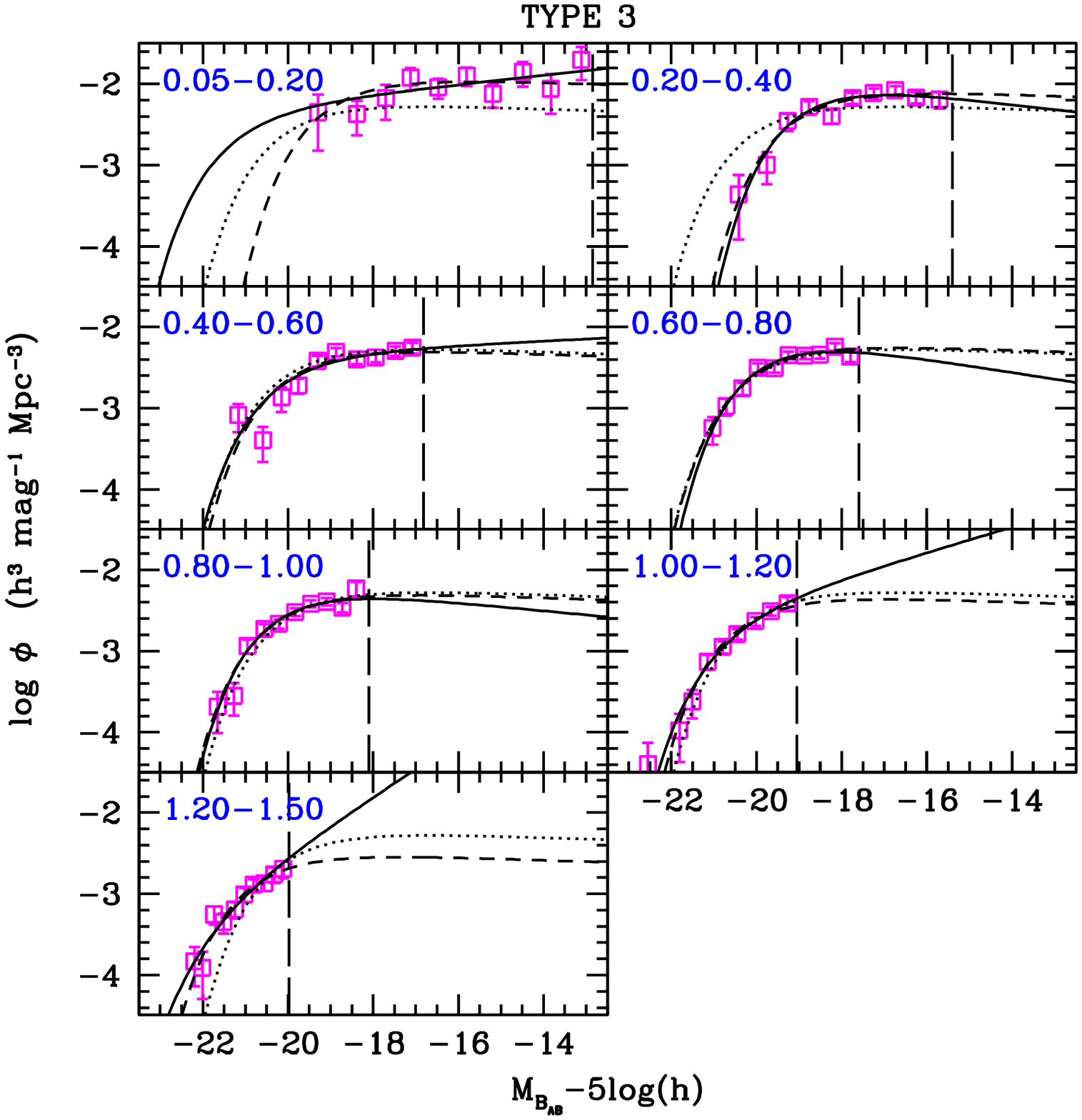}
\caption{
Evolution of the luminosity function in the B-band for type 3 galaxies.
The meaning of the lines and the symbols is the same as in Fig.\ref{LFtype1}.
} 
\label{LFtype3}
\end{figure*}

\begin{figure*}
\centering
\includegraphics[width=\hsize]{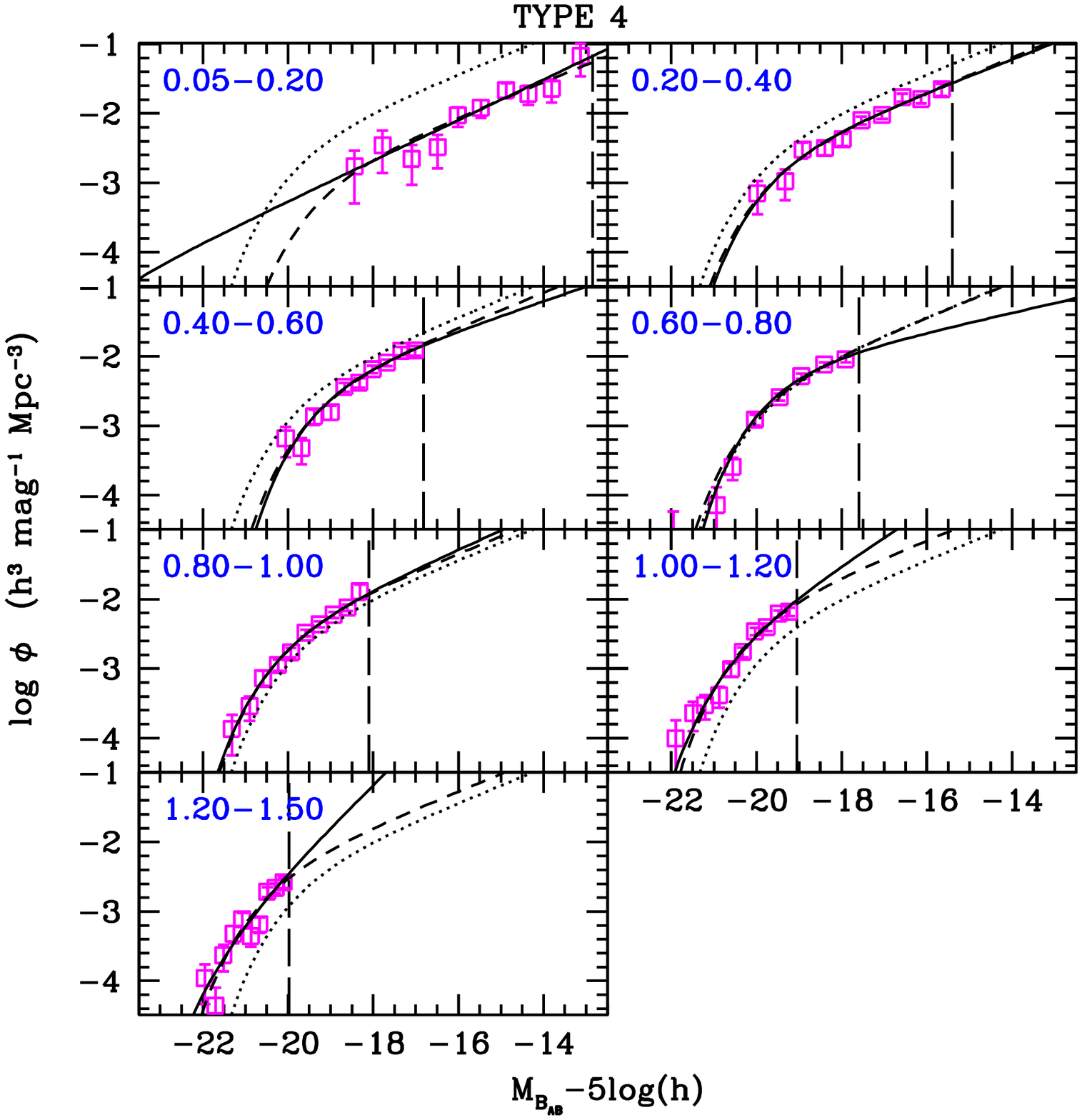}
\caption{
Evolution of the luminosity function in the B-band for type 4 galaxies.
The meaning of the lines and the symbols is the same as in Fig.\ref{LFtype1}.
} 
\label{LFtype4}
\end{figure*}

\begin{table*}
\caption[]{$STY$ parameters (with $1\sigma$ errors) for different galaxy 
types in the rest frame B band} 

\begin{flushleft}

\begin{tabular}{r r r r c l l} \hline
\multicolumn{7}{c}{$\Omega_m$=0.3 \hspace{1cm}                $\Omega_\Lambda$=0.7} \\ \hline
Type & z-bin &  Number$^{(a)}$ &  Number$^{(b)}$ &   $\alpha$ &  $M^*_{AB}-5log(h)$                          & $\phi^*$ ($10^{-3} h^3 Mpc^{-3}$)                    \vspace{0.2cm} \\ \hline
     &       &                 &                 &            &  $\alpha$ free\ \ \ \ \  /\ \ \ \ \  $\alpha$ fixed  & $\alpha$ free\ \ \ /\ \ \  $\alpha$ fixed                      \vspace{0.2cm} \\ \hline
\hline
1    &    0.20-0.40 &    70 &    65 &  -0.04$^{{\rm + 0.28}}_{{\rm - 0.27}}$ 
     & -19.81$^{{\rm + 0.39}}_{{\rm - 0.46}}$   -20.27$^{{\rm + 0.27}}_{{\rm - 0.31}}$ 
     &   5.90$^{{\rm + 0.73}}_{{\rm - 0.73}}$   5.15$^{{\rm + 0.64}}_{{\rm - 0.64}}$ \\ 
     &    0.40-0.60 &   113 &   106 &  -0.40$^{{\rm + 0.20}}_{{\rm - 0.20}}$ 
     & -20.71$^{{\rm + 0.39}}_{{\rm - 0.46}}$   -20.49$^{{\rm + 0.17}}_{{\rm - 0.18}}$ 
     &   2.81$^{{\rm + 0.50}}_{{\rm - 0.58}}$   3.12$^{{\rm + 0.30}}_{{\rm - 0.30}}$ \\ 
     &    0.60-0.80 &   204 &   197 &  -0.22$^{{\rm + 0.17}}_{{\rm - 0.17}}$ 
     & -20.14$^{{\rm + 0.19}}_{{\rm - 0.20}}$   -20.22$^{{\rm + 0.09}}_{{\rm - 0.10}}$ 
     &   3.70$^{{\rm + 0.33}}_{{\rm - 0.43}}$   3.53$^{{\rm + 0.25}}_{{\rm - 0.25}}$ \\ 
     &    0.80-1.00 &   164 &   164 &  -0.01$^{{\rm + 0.25}}_{{\rm - 0.24}}$ 
     & -20.46$^{{\rm + 0.22}}_{{\rm - 0.24}}$   -20.73$^{{\rm + 0.11}}_{{\rm - 0.12}}$ 
     &   2.68$^{{\rm + 0.21}}_{{\rm - 0.21}}$   2.36$^{{\rm + 0.18}}_{{\rm - 0.18}}$ \\ 
     &    1.00-1.20 &   114 &   114 &  -1.23$^{{\rm + 0.34}}_{{\rm - 0.34}}$ 
     & -21.49$^{{\rm + 0.48}}_{{\rm - 0.57}}$   -20.53$^{{\rm + 0.11}}_{{\rm - 0.12}}$ 
     &   0.92$^{{\rm + 0.65}}_{{\rm - 0.56}}$   2.39$^{{\rm + 0.22}}_{{\rm - 0.22}}$ \\ 
     &    0.40-0.90 &   411 &   404 &  -0.29$^{{\rm + 0.10}}_{{\rm - 0.10}}$ & -20.35$^{{\rm + 0.13}}_{{\rm - 0.13}}$ &   3.19$^{{\rm + 0.23}}_{{\rm - 0.26}}$ \\ 
\hline
2    &    0.20-0.40 &   136 &   132 &  -0.67$^{{\rm + 0.13}}_{{\rm - 0.13}}$ 
     & -20.29$^{{\rm + 0.37}}_{{\rm - 0.44}}$   -20.13$^{{\rm + 0.19}}_{{\rm - 0.21}}$ 
     &   5.88$^{{\rm + 1.34}}_{{\rm - 1.33}}$   6.50$^{{\rm + 0.56}}_{{\rm - 0.56}}$ \\ 
     &    0.40-0.60 &   203 &   195 &  -0.50$^{{\rm + 0.15}}_{{\rm - 0.14}}$ 
     & -19.81$^{{\rm + 0.20}}_{{\rm - 0.21}}$   -19.97$^{{\rm + 0.12}}_{{\rm - 0.12}}$ 
     &   4.99$^{{\rm + 0.74}}_{{\rm - 0.79}}$   4.35$^{{\rm + 0.31}}_{{\rm - 0.31}}$ \\ 
     &    0.60-0.80 &   322 &   310 &  -0.57$^{{\rm + 0.13}}_{{\rm - 0.13}}$ 
     & -20.33$^{{\rm + 0.19}}_{{\rm - 0.20}}$   -20.39$^{{\rm + 0.09}}_{{\rm - 0.10}}$ 
     &   4.81$^{{\rm + 0.69}}_{{\rm - 0.74}}$   4.58$^{{\rm + 0.26}}_{{\rm - 0.26}}$ \\ 
     &    0.80-1.00 &   267 &   267 &  -0.60$^{{\rm + 0.20}}_{{\rm - 0.20}}$ 
     & -20.54$^{{\rm + 0.24}}_{{\rm - 0.26}}$   -20.55$^{{\rm + 0.10}}_{{\rm - 0.11}}$ 
     &   3.58$^{{\rm + 0.64}}_{{\rm - 0.74}}$   3.54$^{{\rm + 0.22}}_{{\rm - 0.22}}$ \\ 
     &    1.00-1.20 &   178 &   175 &  -0.76$^{{\rm + 0.34}}_{{\rm - 0.33}}$ 
     & -20.92$^{{\rm + 0.35}}_{{\rm - 0.40}}$   -20.77$^{{\rm + 0.12}}_{{\rm - 0.13}}$  
     &   2.64$^{{\rm + 0.79}}_{{\rm - 0.98}}$   3.01$^{{\rm + 0.23}}_{{\rm - 0.23}}$ \\ 
     &    1.20-1.50 &   103 &   103 &  -1.57$^{{\rm + 0.61}}_{{\rm - 0.62}}$ 
     & -21.65$^{{\rm + 0.62}}_{{\rm - 0.84}}$   -20.82$^{{\rm + 0.13}}_{{\rm - 0.14}}$ 
     &   0.81$^{{\rm + 1.04}}_{{\rm - 0.72}}$   2.19$^{{\rm + 0.22}}_{{\rm - 0.22}}$ \\ 
     &    0.40-0.90 &   677 &   669 &  -0.61$^{{\rm + 0.08}}_{{\rm - 0.08}}$ & -20.25$^{{\rm + 0.12}}_{{\rm - 0.12}}$ &   4.48$^{{\rm + 0.43}}_{{\rm - 0.44}}$ \\ 
\hline
3    &    0.20-0.40 &   341 &   329 &  -0.84$^{{\rm + 0.10}}_{{\rm - 0.10}}$ 
     & -18.92$^{{\rm + 0.19}}_{{\rm - 0.21}}$   -19.14$^{{\rm + 0.12}}_{{\rm - 0.13}}$ 
     &  12.37$^{{\rm + 2.30}}_{{\rm - 2.20}}$   9.82$^{{\rm + 0.54}}_{{\rm - 0.54}}$ \\ 
     &    0.40-0.60 &   451 &   429 &  -1.07$^{{\rm + 0.10}}_{{\rm - 0.10}}$ 
     & -20.28$^{{\rm + 0.24}}_{{\rm - 0.27}}$   -20.04$^{{\rm + 0.11}}_{{\rm - 0.11}}$ 
     &   4.93$^{{\rm + 1.26}}_{{\rm - 1.17}}$   6.31$^{{\rm + 0.30}}_{{\rm - 0.30}}$ \\ 
     &    0.60-0.80 &   626 &   610 &  -0.79$^{{\rm + 0.13}}_{{\rm - 0.13}}$ 
     & -19.86$^{{\rm + 0.17}}_{{\rm - 0.19}}$   -20.10$^{{\rm + 0.09}}_{{\rm - 0.09}}$ 
     &   9.10$^{{\rm + 1.46}}_{{\rm - 1.51}}$   7.11$^{{\rm + 0.29}}_{{\rm - 0.29}}$ \\ 
     &    0.80-1.00 &   534 &   533 &  -0.87$^{{\rm + 0.15}}_{{\rm - 0.15}}$ 
     & -20.23$^{{\rm + 0.18}}_{{\rm - 0.19}}$   -20.33$^{{\rm + 0.08}}_{{\rm - 0.08}}$ 
     &   7.01$^{{\rm + 1.29}}_{{\rm - 1.34}}$   6.27$^{{\rm + 0.27}}_{{\rm - 0.27}}$ \\ 
     &    1.00-1.20 &   292 &   288 &  -1.39$^{{\rm + 0.26}}_{{\rm - 0.26}}$ 
     & -20.82$^{{\rm + 0.31}}_{{\rm - 0.34}}$   -20.38$^{{\rm + 0.10}}_{{\rm - 0.10}}$ 
     &   3.11$^{{\rm + 1.56}}_{{\rm - 1.35}}$   5.57$^{{\rm + 0.33}}_{{\rm - 0.33}}$ \\ 
     &    1.20-1.50 &   208 &   193 &  -1.86$^{{\rm + 0.55}}_{{\rm - 0.59}}$ 
     & -21.87$^{{\rm + 0.77}}_{{\rm - 1.23}}$   -20.81$^{{\rm + 0.12}}_{{\rm - 0.13}}$ 
     &   0.80$^{{\rm + 1.82}}_{{\rm - 0.79}}$   3.67$^{{\rm + 0.27}}_{{\rm - 0.27}}$ \\ 
     &    0.40-0.90 &  1371 &  1349 &  -0.96$^{{\rm + 0.06}}_{{\rm - 0.06}}$ & -20.12$^{{\rm + 0.10}}_{{\rm - 0.11}}$ &   6.79$^{{\rm + 0.72}}_{{\rm - 0.71}}$ \\ 
\hline
4    &    0.20-0.40 &   394 &   380 &  -1.59$^{{\rm + 0.11}}_{{\rm - 0.12}}$ 
     & -19.60$^{{\rm + 0.46}}_{{\rm - 0.58}}$   -19.73$^{{\rm + 0.29}}_{{\rm - 0.33}}$ 
     &   3.05$^{{\rm + 2.09}}_{{\rm - 1.67}}$   2.59$^{{\rm + 0.13}}_{{\rm - 0.13}}$ \\ 
     &    0.40-0.60 &   487 &   449 &  -1.53$^{{\rm + 0.18}}_{{\rm - 0.19}}$ 
     & -19.17$^{{\rm + 0.33}}_{{\rm - 0.39}}$   -19.38$^{{\rm + 0.17}}_{{\rm - 0.18}}$ 
     &   5.57$^{{\rm + 3.14}}_{{\rm - 2.56}}$   4.10$^{{\rm + 0.19}}_{{\rm - 0.19}}$ \\ 
     &    0.60-0.80 &   656 &   622 &  -1.35$^{{\rm + 0.15}}_{{\rm - 0.15}}$ 
     & -19.55$^{{\rm + 0.20}}_{{\rm - 0.21}}$   -19.95$^{{\rm + 0.12}}_{{\rm - 0.12}}$ 
     &   7.72$^{{\rm + 2.33}}_{{\rm - 2.09}}$   4.07$^{{\rm + 0.16}}_{{\rm - 0.16}}$ \\ 
     &    0.80-1.00 &   552 &   552 &  -1.68$^{{\rm + 0.20}}_{{\rm - 0.21}}$ 
     & -20.19$^{{\rm + 0.31}}_{{\rm - 0.36}}$   -20.10$^{{\rm + 0.12}}_{{\rm - 0.12}}$ 
     &   4.06$^{{\rm + 2.44}}_{{\rm - 1.93}}$   4.72$^{{\rm + 0.20}}_{{\rm - 0.20}}$ \\ 
     &    1.00-1.20 &   389 &   373 &  -1.99$^{{\rm + 0.33}}_{{\rm - 0.34}}$ 
     & -20.62$^{{\rm + 0.41}}_{{\rm - 0.52}}$   -20.19$^{{\rm + 0.12}}_{{\rm - 0.12}}$ 
     &   3.19$^{{\rm + 3.49}}_{{\rm - 2.22}}$   6.95$^{{\rm + 0.36}}_{{\rm - 0.36}}$ \\ 
     &    1.20-1.50 &   239 &   188 &  -2.50$^{{\rm + 0.52}}_{{\rm - 0.91}}$ 
     & -21.58$^{{\rm + 0.76}}_{{\rm - 0.40}}$   -20.53$^{{\rm + 0.12}}_{{\rm - 0.12}}$ 
     &   0.52$^{{\rm + 2.06}}_{{\rm - 0.29}}$   4.34$^{{\rm + 0.32}}_{{\rm - 0.32}}$ \\ 
     &    0.40-0.90 &  1442 &  1403 &  -1.62$^{{\rm + 0.08}}_{{\rm - 0.08}}$ & -19.83$^{{\rm + 0.15}}_{{\rm - 0.16}}$ &   4.46$^{{\rm + 1.08}}_{{\rm - 0.95}}$ \\ 
\hline
\multicolumn{7}{l}{(a) Number of galaxies in the redshift bin }\\
\multicolumn{7}{l}{(b) Number of galaxies brighter than the bias limit 
(sample used for $STY$ estimate; see the text for details)}\\

\end{tabular}
\end{flushleft}
\label{parameters}
\end{table*}

\begin{figure*}
\centering
\includegraphics[width=\hsize]{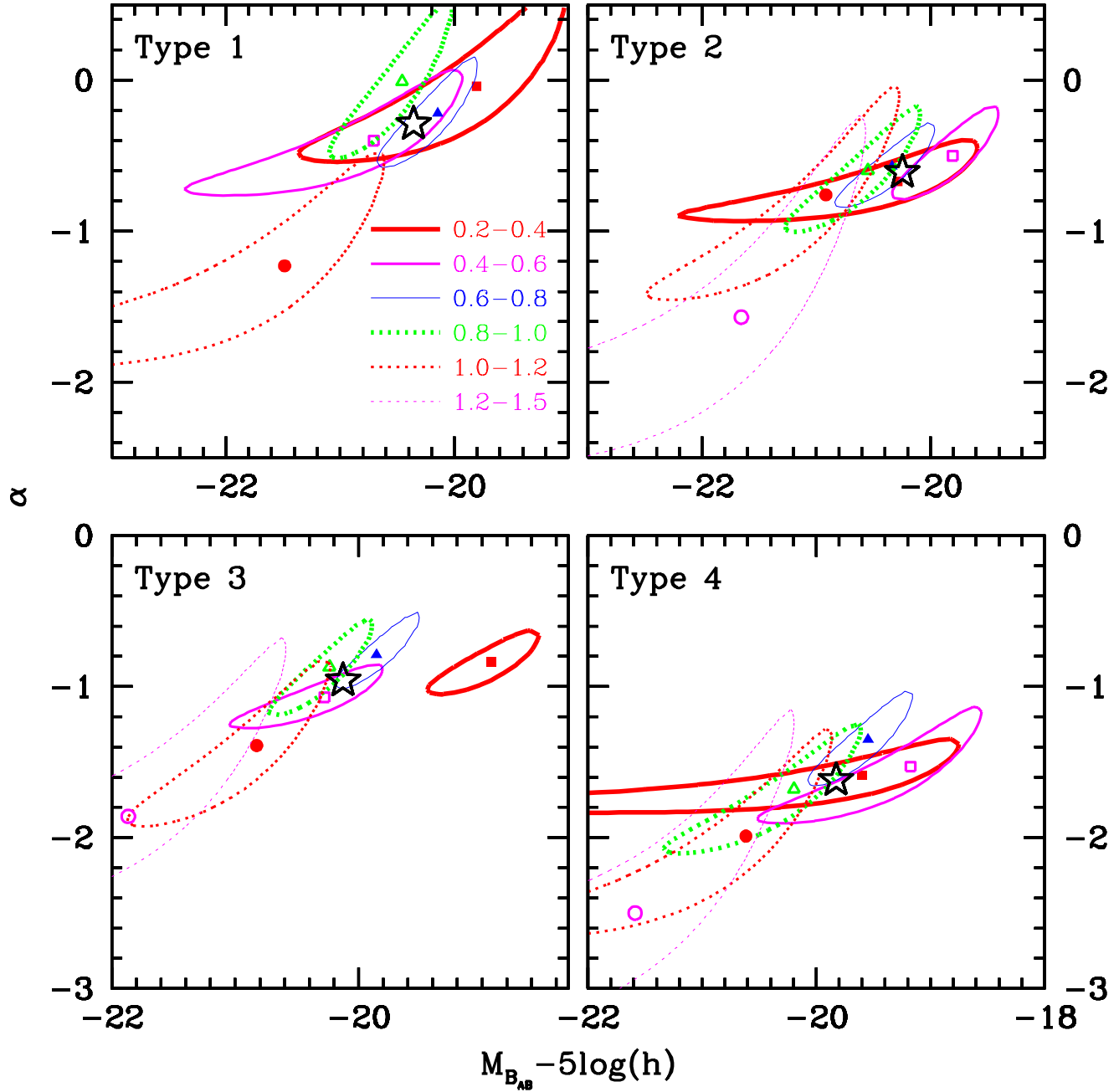}
\caption{
Confidence ellipses at 90\% confidence level for the $\alpha$ and $M^*$ 
parameters of the luminosity functions reported in Fig.\ref{LFtype1}, 
\ref{LFtype2}, \ref{LFtype3} and \ref{LFtype4}. 
Different line types and weights refer to different redshift bins, 
described in the labels; the points indicate the best fit values 
in the redshift bins $[0.2-0.4]$ (filled squares), 
$[0.4-0.6]$ (open squares), $[0.6-0.8]$ (filled triangles),
$[0.8-1.0]$ (open triangles), $[1.0-1.2]$ (filled circles),
$[1.2-1.5]$ (open circles).
The large open star indicates the reference value obtained in the redshift
bin $[0.4-0.9]$. 
} 
\label{ellipses}
\end{figure*}

\begin{figure*}
\centering
\includegraphics[width=\hsize]{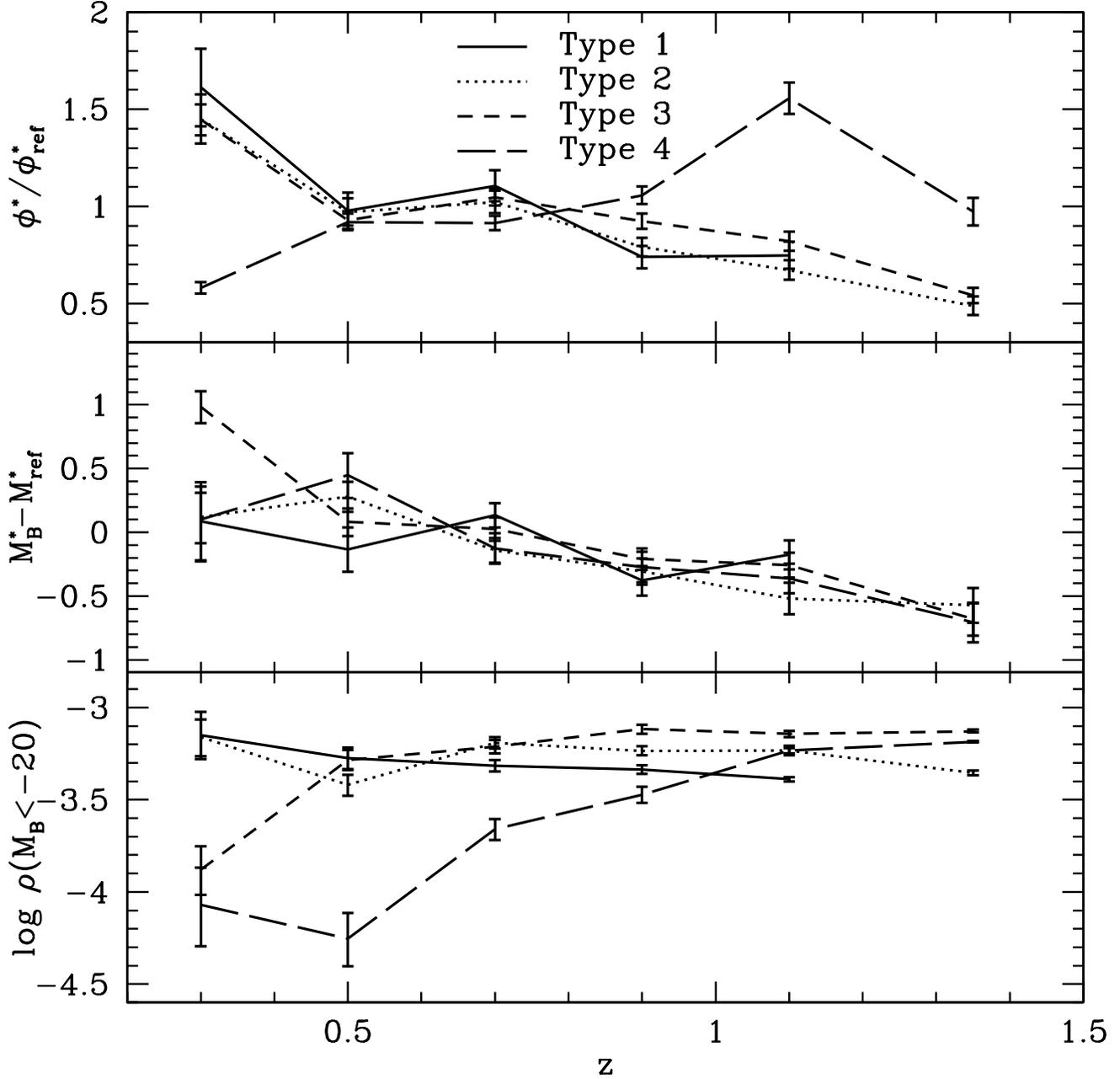}
\caption{
Evolution of the parameters $\phi^*$ (upper panel) and $M^*$ (middle panel)
as a function of the redshift, for different galaxy types. 
In the lowest panel the density of bright ($M_{B_{AB}}-5log(h) < -20$) 
galaxies of different types is shown.
The slope $\alpha$ is fixed to the value derived in the redshift range 
$[0.4-0.9]$; the suffix 'ref' indicates the parameters estimated in the 
redshift range $[0.4-0.9]$. Error bars are at $1\sigma$. 
} 
\label{para_evol}
\end{figure*}

\begin{figure*}
\centering
\includegraphics[width=0.45\hsize]{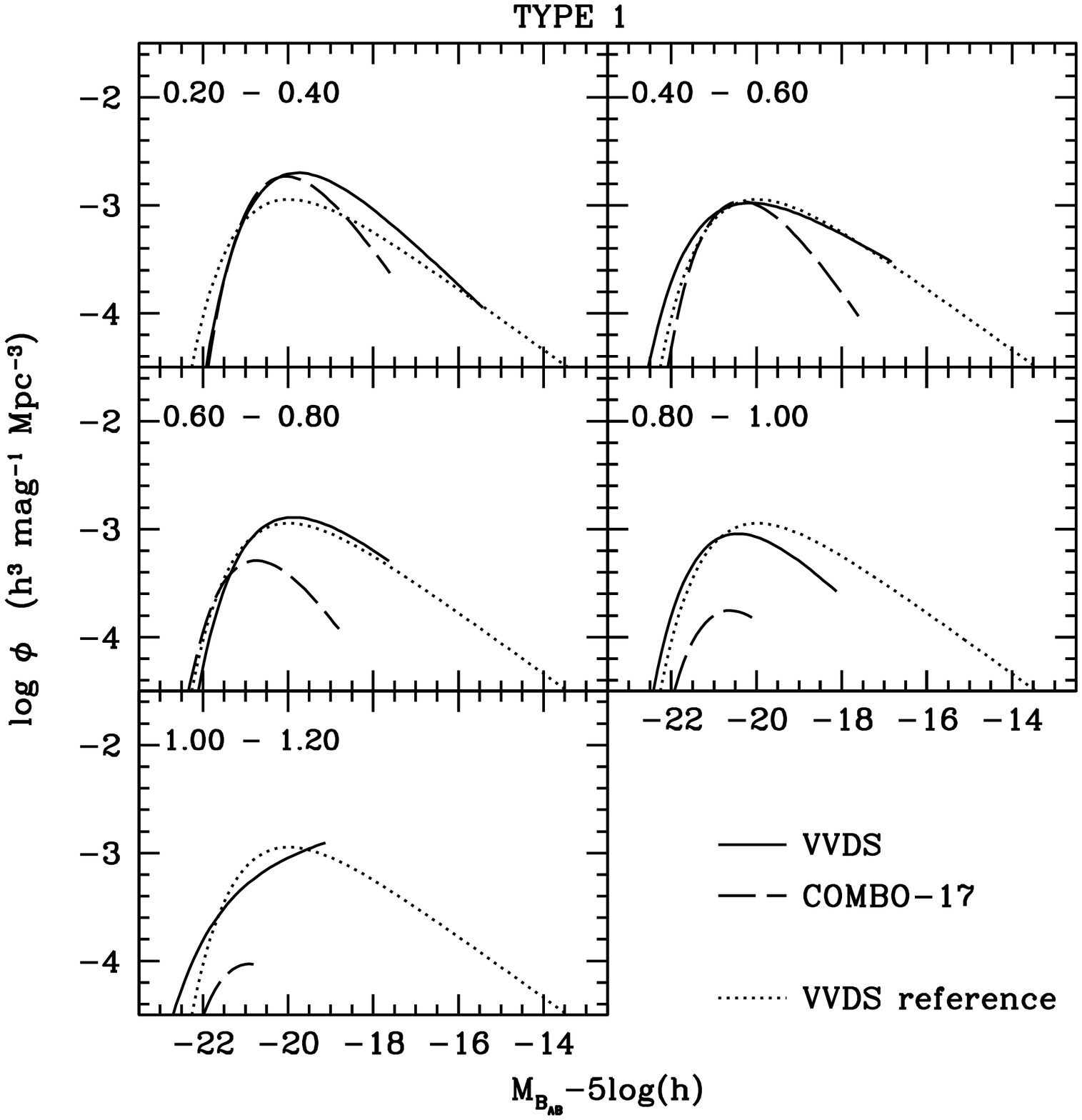}
\includegraphics[width=0.45\hsize]{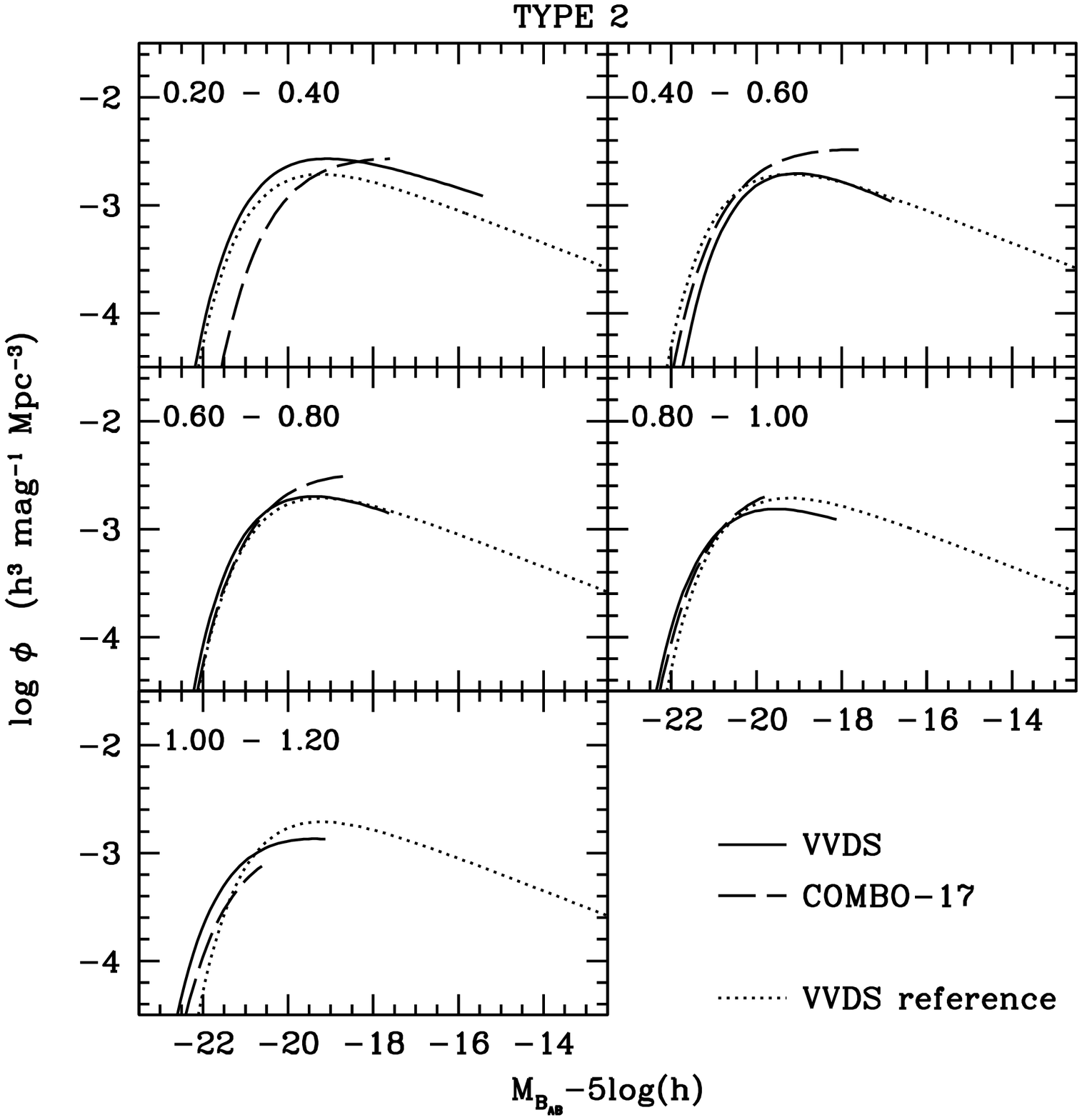}
\includegraphics[width=0.45\hsize]{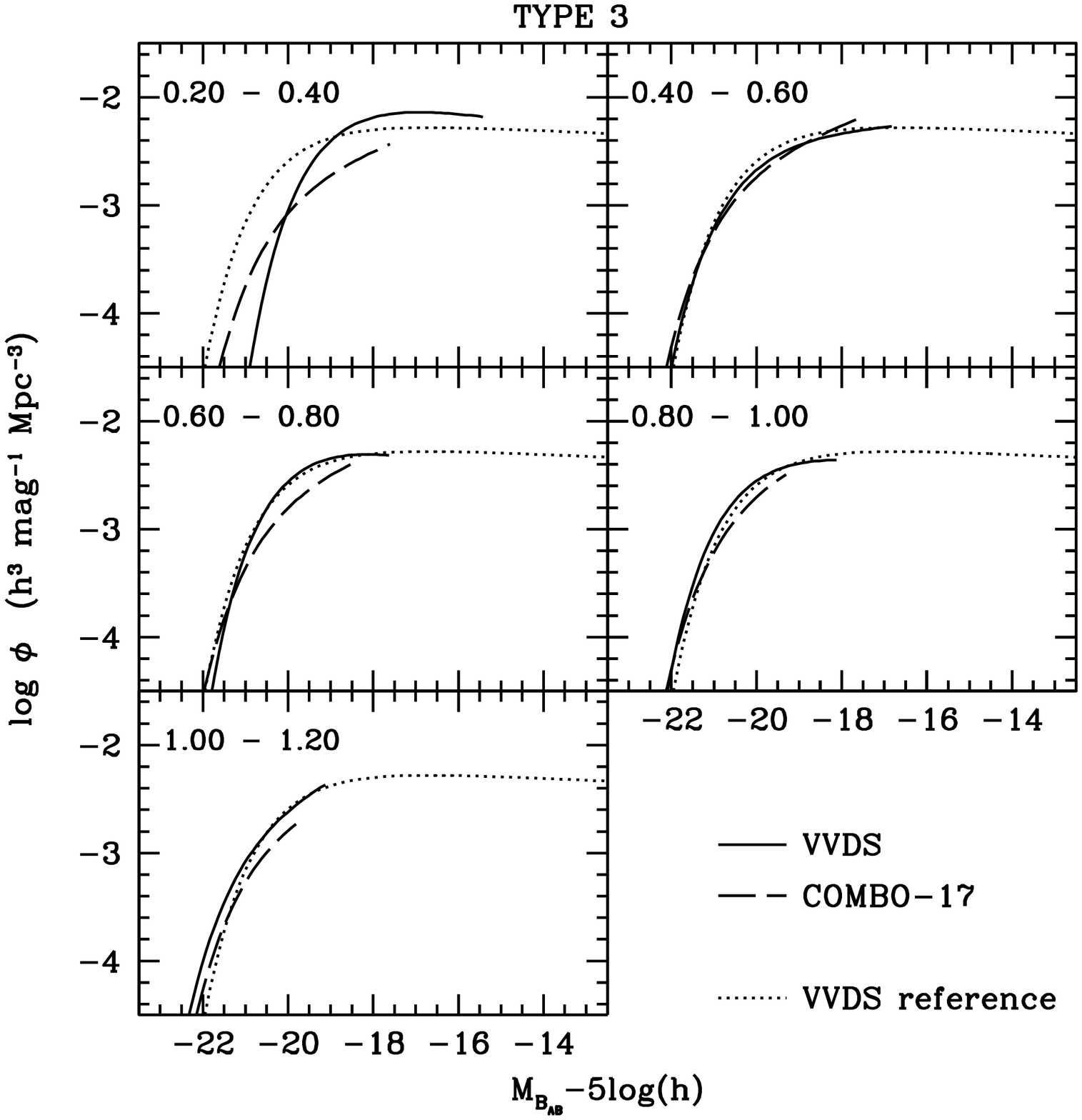}
\includegraphics[width=0.45\hsize]{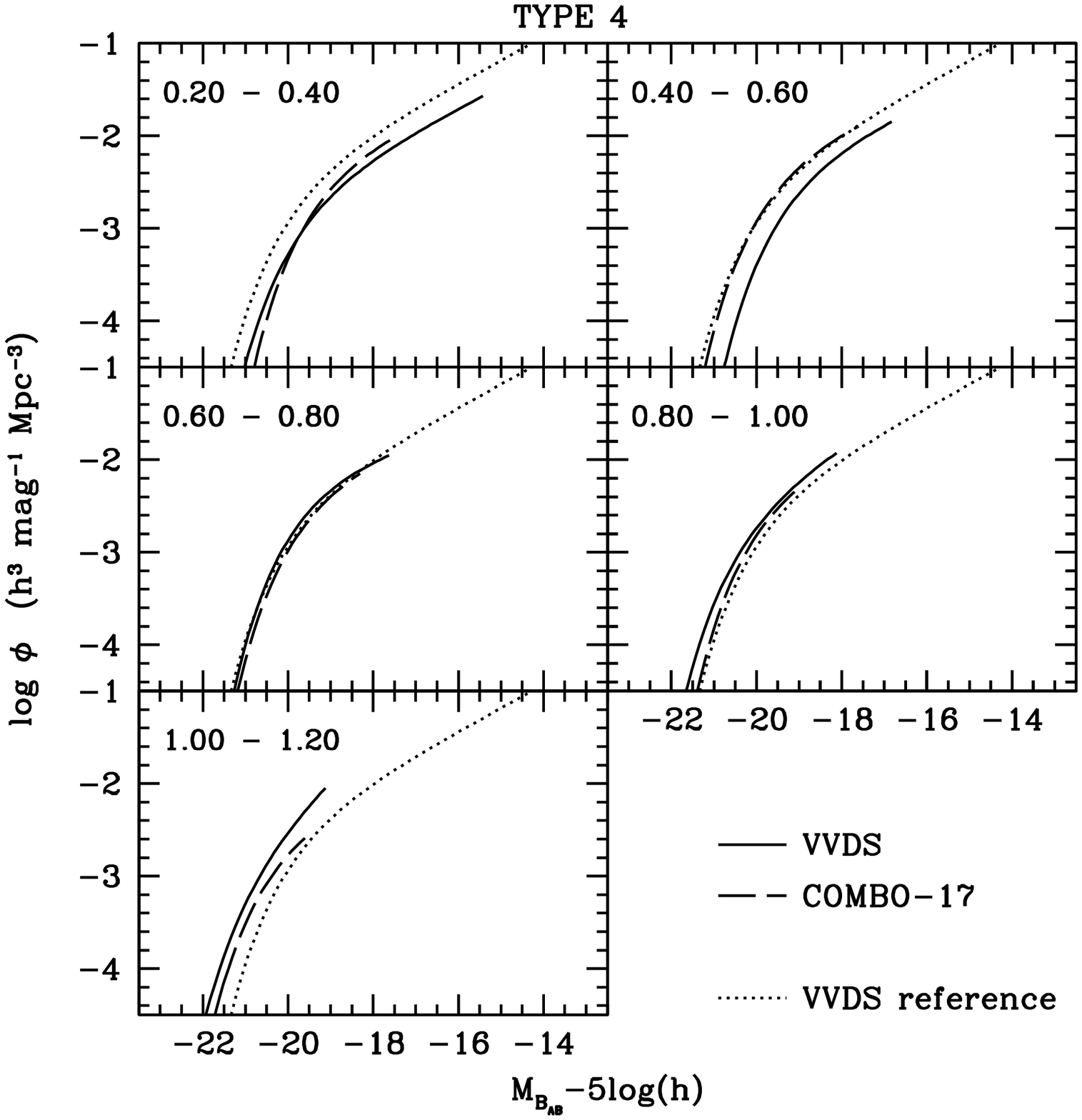}
\caption{
Comparison between VVDS and COMBO-17 luminosity functions, in various
redshift bins (indicated in the label in each panel) and for various types.
Upper panels: type 1 (left) and type 2 (right) galaxies. Lower panels:
type 3 (left) and type 4 (right) galaxies.
Solid lines: VVDS estimate. Dotted lines: VVDS estimate in the redshift
range $[0.4-0.9]$, plotted as a reference. Dashed line: COMBO-17 estimate
from Wolf et al. (\cite{wolf03}).
} 
\label{LFcombo}
\end{figure*}

\section{Evolution with redshift of the luminosity functions by type}

We derived luminosity functions for each type in redshift bins in the 
U, B, V, R and I rest frame bands. 
Given our multicolour coverage and the explored redshift range, the 
estimate of the absolute magnitudes in the U and B rest frame bands 
are those which require less extrapolations (see Appendix A and 
Figure A.1 in Ilbert et al. \cite{vvdsLF}). Therefore, to limit the
number of figures in the paper, we show the results in the B rest frame
band. 
\\
Figures \ref{LFtype1}, \ref{LFtype2}, \ref{LFtype3} and \ref{LFtype4} show 
the luminosity function for type 1, 2, 3 and 4 galaxies in
redshift bins, obtained with $C^+$ and $STY$ methods.
The luminosity functions derived with the other two methods ($1/V_{max}$ 
and $SWML$) are consistent with those shown in the figures, but are not 
drawn for clarity. 
The dotted line in each panel represents the fit derived in the redshift range 
$[0.4-0.9]$ (see previous section), while the dashed line is the estimate 
derived by fixing the slope $\alpha$ to the value obtained in the range 
$[0.4-0.9]$.
\\
In Table \ref{parameters} we report the Schechter parameters, with
their $1\sigma$ errors, estimated
for the various redshift bins, from $z=0.2$ to $z=1.5$ for each type; 
as a reference, in the last line we give the parameters derived in the 
redshift bin $[0.4-0.9]$.
\\
We do not show the results for bins where the number of objects is
too small (less than $\sim 30$) to constrain the parameters of the 
luminosity function (these are the bin $[0.05-0.2]$ for type 1 and 
2 and the bin $[1.2-1.5]$ for type 1).
Given the bright magnitude limit of the survey (I$_{AB} \ge 17.5$) and the 
small sampled volume, bright galaxies are not sampled in the redshift 
bin $[0.05-0.2]$  and therefore we can not constrain the $M^*$ 
parameter even for type 3 and 4 galaxies, where the number of objects is 
relatively high ($\sim 80$). 
For this reason we show in the figures the luminosity function 
estimates in this bin, but we do not report the $STY$ parameters for this 
redshift range in Table \ref{parameters}. 
\\
In Fig.\ref{ellipses} we show the confidence ellipses of the parameters
$\alpha$ and $M^*$ in different redshift bins for the different types.
From this figure it is possible to see that, within each type, the estimated 
slopes $\alpha$ in the various redshift bins are always consistent (within 
90\% confidence level) with each other and with the value derived in the 
redshift range $[0.4-0.9]$. Therefore, there is no evidence of a significant 
change with redshift of the luminosity function slope within each galaxy type.
Note also that the uncertainties on the slope estimates become quite
large for $z>1$: this is due to the fact that, even with the faint limit
(I$_{AB} \le 24$) of this survey, the number of galaxies fainter than
$M^*$ is too low to well constrain the slope. 
\\
In each panel of Figures \ref{LFtype1}, \ref{LFtype2}, \ref{LFtype3} and 
\ref{LFtype4} we draw, as a reference, the luminosity function derived
in the redshift bin $[0.4-0.9]$ (dotted line). Comparing this curve with
the estimates of the luminosity function in the different redshift bins 
an evolution can be seen, strongly depending on the galaxy type.
This evolution is particularly evident for type 4 galaxies: going from 
low to high redshift there is an almost continuous brightening of $M^*$ and   
at fixed luminosity the density of these galaxies was much higher in the
past.
\\
The observed evolution of the luminosity function could be due to an evolution
in luminosity and/or in density or both. 
One of the advantages of the $STY$ method is that of allowing to derive 
the $\alpha$ and $M^*$ parameters independently from $\phi^*$, which is not 
possible when one directly fits a Schechter function on the $1/V_{max}$ points.
\\
Given the fact that we found that $\alpha$ is consistent with being
constant for each type, we can fix it at the reference value derived
in the redshift range $[0.4-0.9]$ and then study the variations 
of the parameters $M^*$ and $\phi^*$ as a function of the redshift
(see upper and middle panels of Fig.\ref{para_evol}). 
These estimates are reported in Table \ref{parameters}. 
\\
From Fig.\ref{para_evol} we can see a mild evolution of $M^*$ 
from the lowest to the highest redshift bin for each type.
In particular, this brightening ranges from $\simlt 0.5$ mag for early type
galaxies to $\sim 1$ mag for the latest type galaxies.
The only exception with respect to the general trend is that of type 3 objects 
in the bin $[0.2-0.4]$, for which the best fit value of the $M^*$ parameter 
is significantly fainter than expexted. The reason for this discontinuity
in $M^*$ for type 3 galaxies at low redshift is not clear.
\\
On the contrary, the $\phi^*$ parameter shows a very different behaviour
for type 1, 2 and 3 galaxies with respect to type 4 galaxies.
The first three types show a rapid decrease of $\phi^*$ at low redshifts
(between $z\sim 0.3$ and $z\sim 0.5$), then $\phi^*$ remains roughly constant 
up to $z\sim 0.9$ and finally slowly decreases up to $z=1.5$. 
\\
Type 4 objects, on the contrary, show an increase in $\phi^*$ at low 
redshift, then $\phi^*$ is nearly constant up to $z\sim 0.8$ and shows a rapid 
increase of a factor $\sim 2$ at $z=1.1$. Then there seems to be a decrease
from $z=1.1$ to $z=1.3$. However, this decrease can likely be a spurious effect,
due to the fact that in this bin the estimated $M^*$ is very close 
to the bias limit (see Sect. 4). If, for example, we fix $M^*$ to the 
value obtained in the previous redshift bin $[1.0-1.2]$, we derive 
a significantly higher value for $\phi^*$, i.e. 
$\phi^* = 5.83 \times 10^{-3} h^3 Mpc^{-3}$
(corresponding to $\phi^* / \phi^*_{ref} = 1.31$). 
Therefore the last $\phi^*$ value for type 4 galaxies is likely to be 
a lower limit of the true density value. 
\\
This analysis of the different trends with redshift of the $\phi^*$ parameter
indicates that the importance of type 4 galaxies is increasing
with redshift. However, since $M^*$ is changing with redshift (see above),
the trends of $\phi^*$ can not be immediately interpreted in terms of density
at a given absolute magnitude. For this reason, we have computed the
density of bright galaxies as a function of redshift. 
We integrated the best fit luminosity function down to 
$M_{B_{AB}}-5log(h) < -20$. This limit approximately corresponds to the
faintest galaxies which are visible in the whole redshift range.
In the lowest panel of Fig.\ref{para_evol} we plot the density of 
bright galaxies of each type as a function of redshift. 
\\
The main results shown in this plot can be summarized in the following
way:
\\
a. the density of bright early type galaxies (type 1) decreases with
increasing redshift, however this decrease is rather modest, 
being of the order of $\sim 40\%$ from $z\sim 0.3$ to $z\sim 1.1$;  
\\
b. the density of bright late type galaxies (type 4) is instead significantly
increasing, by a factor $\sim 6.6$ from $z\sim 0.3$ to $z\sim 1.3$.
\\
The behaviour of type 4 galaxies is also responsible of the evolution of
the global luminosity function measured by Ilbert et al. (\cite{vvdsLF}).
In fact, the increasing number of both faint and bright type 4 galaxies
leads to the steepening of the global LF (due to the very steep slope of
type 4 LF) and to the brightening of $M^*$ (due to the increasing 
fraction of bright blue objects). 
This fact has been directly checked summing the LF of all types and 
comparing the result with the global LF estimate.

\section{Comparison with previous literature results}

\begin{figure}
\centering
\includegraphics[width=\hsize]{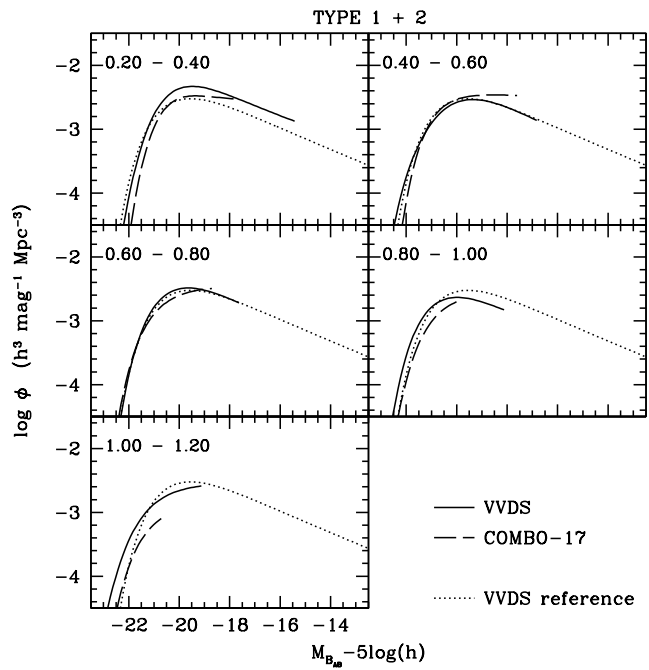}
\caption{
Comparison between VVDS and COMBO-17 luminosity functions, in various
redshift bins (indicated in the label in each panel) for the sum 
of types 1 and 2.
The meaning of the symbols is the same as in Fig.\ref{LFcombo}. 
}
\label{LFcombo12}
\end{figure}

Although various estimates of the luminosity function by galaxy type are
available in the literature, a quantitative comparison of our results with
previous analyses is not straightforward, because of the different
classification schemes adopted in the various surveys, the different
numbers of galaxy types, 
the different redshift ranges and selection criteria.
\\
The strong density evolution of late type galaxies we find in the redshift
range $[0.2-1.5]$ extends to significantly higher redshift the
results found by de~Lapparent et al. (\cite{delapparent04}) for their
latest type at $z\sim 0.5$. 
\\
Among the other surveys, CNOC-2 (Lin et al. \cite{lin99}) and
COMBO-17 (Wolf et al. \cite{wolf03}) adopted a classification scheme 
somewhat similar to ours.
Lin et al. (\cite{lin99}) divided their sample of galaxies with $z<0.55$
in three classes (early, intermediate and late type) using CWW templates. 
For early type galaxies they found a positive luminosity evolution, 
which is nearly compensated by a negative density evolution. 
On the contrary, for late type galaxies they found a strong
positive density evolution, with nearly no luminosity evolution.
We see at higher redshift the same trend in density, but our evolution 
in $M^*$ is contained within one magnitude for each type. 
Wolf et al. (\cite{wolf03}) used a sample of $\sim 25000$ galaxies with
photometric redshifts, applying a classification scheme in four classes
similar to ours but using the Kinney et al. (\cite{kinney96})
templates instead of the CWW templates.  
\\
In Figure \ref{LFcombo} we compare our results (solid lines)
with those from COMBO-17 (dashed lines). 
Note that, since our data extend always to fainter absolute magnitudes, 
in particular for type 1 galaxies, our estimates of the faint end slope 
are likely to be better determined: these estimates are derived
in each redshift bin, while the COMBO-17 slopes are
fixed to the value determined in the redshift range [0.2 - 0.4].
This figure shows that there are significant differences in both shapes 
and evolution of the LF estimates.
The slope of the COMBO-17 LF is flatter than ours for type 1 galaxies, 
while it is steeper for types 2 and 3 (at least up to $z=1.0$). 
The most significant difference between the two surveys regards the 
evolution of type 1 galaxies, for which we do not have any evidence
of the very strong density decrease with increasing redshift 
present in COMBO-17 data.
The reason for this difference is unclear: it could be due to the use of
different templates in the definition of the galaxy types or to a degeneracy
between photometric redshift and classification, which might affect the
COMBO-17 data. 
Bell et al. (\cite{bell04}) explained the strong negative evolution of type 1
galaxies as a consequence of the blueing with increasing redshift of
elliptical galaxies with respect to the template used for classification.
In this way, at increasing redshift an increasing number of ``ellipticals'' 
would be assigned a later type, therefore producing the detected 
density decrease observed in COMBO-17 data.
However, as already mentioned in sect.3, we verified that this effect 
does not affect our classification scheme, at least up to $z\sim 1$ and
for simple stellar populations with $z_{form}>2$.
\\ 
As a test of this possible effect, we compared the LF obtained
by adding together type 1 and 2 objects. In this way
we can check whether the differences between  VVDS and COMBO-17 
LF of type 1 galaxies is due to the fact that a significant fraction
of high redshift type 1 galaxies in COMBO-17 are classified as type 2 
galaxies. This comparison is shown in Fig.\ref{LFcombo12}.
The discrepancy between VVDS and COMBO-17 LF is now reduced,  
but there are still significant differences both in slope 
(being the COMBO-17 LF steeper than that of VVDS) 
and in normalization, especially in the highest redshift bin,
where the VVDS LF is more than a factor 2 higher than the COMBO-17 LF. 

\section{Conclusions}

In this paper we studied the evolution of the luminosity function of
different galaxy types up to $z=1.5$, using 7713 spectra with $17.5 \le$ 
I$_{AB}\le 24$ from the first epoch VVDS deep sample.
\\
The VVDS data allow for the first time to study with excellent statistical
accuracy the evolution of the luminosity functions by galaxy type from 
relatively low redshift up to $z=1.5$ from a single purely 
magnitude selected spectroscopic sample.
The faint limiting magnitude of the VVDS sample
allows to measure the slope of the faint end of the luminosity
function with unprecedented accuracy up to $z\sim 1.2$.
The use of spectroscopic redshifts implies low
``catastrophic'' failure rate compared to the photometric redshifts
and therefore rare populations are sampled with a better accuracy,
like in the bright end on the luminosity function.
Moreover, the use of spectroscopic redshifts allow to 
classify galaxies avoiding a possible degeneracy between photometric 
redshift and classification. 
\\
VVDS galaxies were classified in four spectral classes using their colours and
redshift, from early type to irregular galaxies, and luminosity functions 
were derived for each type in redshift bins, from $z=0.05$ to $z=1.5$, in the
U, B, V, R and I rest frame bands.
\\
We find a significant strong steepening of the luminosity function going
from early to late types: in all bands the power law slope steepens by
$\Delta\alpha \sim 1.3-1.5$ going from type 1 to type 4.
Moreover, the $M^*$ parameter of the Schechter function is significantly 
fainter for late type galaxies. As expected, this difference increases
in the redder bands, reaching $\sim 1.4$ mag in the I band. 
\\
Studying the variations with redshift of the luminosity function for each
type, we find that there is no evidence of a significant change of the
slope, while we find a brightening of $M^*$ with increasing redshift, 
ranging from $\simlt 0.5$ mag for early type galaxies to $\sim 1$ mag 
for the latest type galaxies.
We also find a strong evolution in the normalization of the luminosity function
of latest type galaxies, with an increase of more than a factor $2$ in the
$\phi^*$ parameter going from $z\sim 0.3$ to $z\sim 1.3$.
The density of bright ($M_{B_{AB}}-5log(h) < -20$) galaxies shows a modest decrease
($\sim 40\%$) for early type objects from $z\sim 0.3$ to $z\sim 1.1$; on the
contrary, the number of bright late type galaxies increases of a factor 
$\sim 6.6$ from $z\sim 0.3$ to $z\sim 1.3$.  
\\
Our results indicate that the importance of type 4 galaxies is increasing
with redshift, with an important contribution of both bright and faint
blue objects. This fact is also largely responsible of the evolution of 
the global luminosity function measured by Ilbert et al. (\cite{vvdsLF}), 
which shows a brightening of $M^*$ and a steepening of $\alpha$ with
increasing redshift. Moreover, the increasing
contribution of blue galaxies has been seen in the evolution of the GALEX-VVDS 
luminosity function at 1500\AA (Arnouts et al. \cite{galexLF}). 
We are therefore pinpointing that the  galaxies responsible for
most of the evolution quoted in the literature belong to the population of the
latest spectral type. The epoch at which a transition between a Universe
dominated by late type galaxies and a Universe dominated by old massive
objects occurs is at a redshift of $z\sim 0.7-0.8$.  
\\
The fact that type 1 galaxies show only a mild evolution both in luminosity
(positive) and in density (negative) is consistent with the fact that
most of the objects in this class are old ($z_{form} > 2$, see sect. 3) 
galaxies, experiencing only a passive evolution in the explored redshift range.
More intriguing is the density evolution of type 4 galaxies, which corresponds
to an increasing number of bright star forming galaxies towards high redshift
which could be connected to various populations of high redshift objects seen in
multiwavelength surveys.

\begin{acknowledgements}
This research has been developed within the framework of the VVDS
consortium.\\
This work has been partially supported by the
CNRS-INSU and its Programme National de Cosmologie (France),
and by Italian Ministry (MIUR) grants
COFIN2000 (MM02037133) and COFIN2003 (num.2003020150).\\
The VIMOS VLT observations have been carried out on guaranteed
time (GTO) allocated by the European Southern Observatory (ESO)
to the VIRMOS consortium, under a contractual agreement between the
Centre National de la Recherche Scientifique of France, heading
a consortium of French and Italian institutes, and ESO,
to design, manufacture and test the VIMOS instrument.
\end{acknowledgements}



\end{document}